\documentclass[pdftex,twocolumn,epjc3]{svjour3}          % twocolumn

\RequirePackage[T1]{fontenc}

\smartqed  % flush right qed marks, e.g. at end of proof

\RequirePackage{graphicx}
\RequirePackage{mathptmx}      % use Times fonts if available on your TeX system
\RequirePackage{flushend}
\RequirePackage[numbers,sort&compress]{natbib}
\RequirePackage[colorlinks,citecolor=blue,urlcolor=blue,linkcolor=blue]{hyperref}
\usepackage{subfig}
\usepackage{microtype}
\usepackage{amsmath}
\usepackage{amssymb}
\usepackage{soul}
\usepackage{booktabs}
\usepackage{float}

\journalname{Eur. Phys. J. C}

\usepackage{soul}

\begin{document}

\title{Probing Low-Luminosity Gamma-Ray Emission from  SNR G296.5+10.0 and CCO 1E 1207.4-5209 with CTAO}

\author{Luana N. Padilha\thanksref{e1,addr1}
\and
Rubens Jr. Costa\thanksref{e2,addr2}
\and
Rita C. dos Anjos\thanksref{e3,addr1,addr2,addr3,addr4,addr5}
\and
Jaziel G. Coelho\thanksref{e4,addr2,addr5}
}

\thankstext{e1}{e-mail:luana.natalie.padilha@uel.br}
\thankstext{e2}{e-mail:rubensp@utfpr.edu.br}
\thankstext{e3}{e-mail:ritacassia@ufpr.br}
\thankstext{e4}{e-mail:jaziel.coelho@ufes.br}

\institute{Department of Physics, Universidade Estadual de Londrina (UEL), Rodovia Celso Garcia Cid Km 380, 86057-970 Londrina, PR, Brazil \label{addr1} 
\and 
Postgraduate Program in Physics and Astronomy, Universidade Tecnológica Federal do Parana (UTFPR), Av. Sete de Setembro, 3165, 80230-901, Curitiba, PR, Brazil \label{addr2} 
\and 
Department of Engineering and Exact Sciences, Universidade Federal do Parana (UFPR), Pioneiro, 2153, 85950-000 Palotina, PR, Brazil \label{addr3} 
\and 
Postgraduate Program in Applied Physics, Universidade Federal da Integração Latino-Americana  (Unila), 85867-670, Foz do Igua\c{c}u, PR, Brazil \label{addr4}  
\and 
Astrophysics and Cosmology Center, Universidade Federal do Esp\'irito Santo (Cosmo-Ufes), 29075--910, Vit\'oria, ES, Brazil \label{addr5}         
}

\date{ Accepted: 30/04/2026}

\maketitle

\sloppy
\begin{abstract}
The acceleration mechanisms of cosmic rays (CR) in supernova remnants (SNRs) and their associated compact central objects (CCOs) remain an open question in high-energy astrophysics. In this study, we perform a modeling of CR transport and gamma-ray emission from SNR G296.5+10.0 and its CCO 1E~1207.4-5209, using the latest public release of the \texttt{GALPROP} code (v57) and focusing, in particular, on the contribution from the CCO. Our simulations predict the contribution of CR from this source to the Galactic flux, accounting for energy losses and particle interaction processes. We find that, under time-evolving scenarios, the environment around SNR G296.5+10.0 and 1E~1207.4-5209 is suitable for CR acceleration and gamma-ray production. The analysis distinguishes between gamma rays produced by hadronic interactions in SNR G296.5+10.0 and by leptonic processes in CCO 1E~1207.4-5209, revealing that each mechanism dominates in different energy bands. We show that the Cherenkov Telescope Array Observatory (CTAO) can detect this emission with a significance of $5\sigma$ after 50~h of exposure, providing the first constraints on particle acceleration in this unique CCO-SNR system. These findings suggest that CCOs may be efficient electron accelerators, even in the absence of pulsar wind nebulae, and emphasize the critical role of next-generation observatories such as CTAO in unraveling CR acceleration processes in low-luminosity SNR-CCO systems.

\end{abstract}
\fussy

\section{Introduction}
\label{sec:intro}

The origin and acceleration mechanisms of energetic particles in the Universe remain among the most pressing questions in high-energy astrophysics. Supernova remnants (SNRs) are widely recognized as key contributors to the acceleration of cosmic rays (CR) and play a crucial role in shaping the observed CR spectrum. Among the compact objects associated with SNRs, central compact objects (CCOs) stand out due to their unique properties, such as intense thermal X-ray emission and the absence of significant radio or gamma-ray counterparts \cite[see][]{Hillas2004,Ptitsyna2010b,harding2013neutron,2004IAUS..218..239P,2017JPhCS.932a2006D}. These properties make CCO-SNR systems particularly relevant for studying particle acceleration in environments with strong radiative and gravitational fields, offering insights into the physics and evolution of young neutron stars (NSs) and their interaction with supernova ejecta \cite{2021A&A...651A..40M}. 

The relationship between CCOs and SNRs is closely tied to the nature of their progenitor explosion, influencing their morphology and interactions with surrounding matter and magnetic fields \cite{lopez2018morphologies}. CCOs located at the geometric centers of SNRs typically emit thermal X-rays in the hundreds of electronvolts (eV) range, with luminosities between $10^{33}$ and $10^{34}\,\mathrm{erg\,s^{-1}}$ \cite[see, e.g.,][]{2010ApJ...709..436H,2004IAUS..218..239P}. Due to their elusive nature, only ten CCOs have been officially cataloged, although many more likely remain undetected \cite{2021A&A...651A..40M}. Their rotational periods and spin-down rates suggest that CCOs have unusually weak magnetic fields, ranging from $10^{10}$ to $10^{11}\,\mathrm{G}$, probably as a result of magnetic dipole braking and consistent with their relatively young age \cite{2010ApJ...709..436H}.

SNR G296.5+10.0, together with its CCO 1E~1207.4-5209, provides a clear example of a CCO-SNR system, illustrating the unique physical conditions that can influence particle acceleration and magnetic field evolution. The SNR G296.5+10.0 exemplifies a bilateral remnant, displaying a barrel-shaped morphology in radio observations, with two bright opposing limbs aligned nearly perpendicular to the Galactic plane \cite{1987A&A...183..118K,2024MNRAS.528.2095E}, providing a compelling setting for investigating CR acceleration and gamma-ray production. The presence of CCO 1E~1207.4-5209 near the geometric center of the SNR further enhances its scientific appeal. The association between 1E~1207.4-5209 and SNR G296.5+10.0 offers a rare opportunity to explore the relationship between young NSs and their host remnants, particularly in low-luminosity environments where particle acceleration processes can be masked or suppressed by environmental effects.

1E~1207.4-5209 stands out from other CCOs mainly due to the discrepancy between its estimated age of $\sim\!10^8$~yr (according to the ATNF\footnote{Australia Telescope National Facility: \url{https://www.atnf.csiro.au/research/pulsar/psrcat/}} catalog) and the age of its associated SNR, which is about $\sim\!10^4$~yr. In addition, its magnetic field has been directly measured, unlike most objects in this class and many pulsars, whose magnetic field strength is typically inferred. It is also noteworthy that 1E~1207.4-5209 was the first CCO detected with pulsed X-ray emission, exhibiting a highly stable rotation period with minimal magnetic activity \cite{1987A&A...183..118K,2024MNRAS.528.2095E,1987AuJPh..40..815K}. Although it shares typical CCO characteristics, such as strong thermal X-ray emission and a relatively weak magnetic field, these features make 1E~1207.4-5209 an ideal testbed for particle acceleration models in low magnetic field environments. Although previous studies \cite{2004A&A...418..625D} have characterized its thermal properties, its gamma-ray potential remains largely unexplored. 

The region of the SNR G296.5+10.0 has been extensively studied as a high-energy gamma-ray emitter, particularly with respect to its lepto-hadronic spectral energy distribution (SED), covering synchrotron emission, bremsstrahlung, inverse Compton (IC) scattering, and proton--proton interactions \cite{2024MNRAS.528.2095E,2013MNRAS.434.2202A,2021ApJ...910...78Z}. Given the low luminosity of the region, much remains to be explored, particularly regarding potential nuclear injection processes. Further observations will be crucial for unveiling its true capabilities. The Cherenkov Telescope Array Observatory (CTAO), poised to become the leading ground-based gamma-ray observatory of the next decade \cite{2019EPJWC.20901038C,2019scta.book.....C}, offers promising opportunities to detect gamma-ray emission from this region, potentially advancing our understanding of it. Detecting gamma rays would also provide information on the interplay between leptonic and hadronic processes \cite{2025BrJPh..55...60S}.

In light of this, we investigate the gamma-ray emission and CR propagation from SNR G296.5+10.0 \cite{2011A&A...525A.106D}, focusing on the CCO 1E~1207.4-5209. Using the latest version of the \texttt{GALPROP} 
\footnote{\url{https://galprop.stanford.edu/}} code, which includes CR injection and transport, we estimate the contribution of CR from this source, explicitly accounting for energy losses, diffusion delays, and particle interactions during propagation \cite{1998ApJ...509..212S,2002ApJ...565..280M,2007ARNPS..57..285S,galprop}. Our analysis also evaluates the CTAO's capability to detect gamma-ray emission from this region, emphasizing a low-luminosity scenario in which the interplay between particle acceleration and environmental factors remains poorly understood.

Our research is part of the multimessenger effort to trace CR origins, which uses both the cosmic rays themselves and the gamma rays they produce. In particular, accelerated electrons and positrons contribute through synchrotron emission, bremsstrahlung, and inverse Compton scattering \cite{2015EPJWC.10500003B,2020PhR...872....1B}. The correlation between CR propagation and the integrated flux of secondary gamma rays allows modeling of source luminosities both within and beyond our Galaxy \cite{Supanitsky_2013,Anjos_2014,Sasse_2021,2021JCAP10023D,coelho2022updated,2025ApJ...994...31S}. In this way, the detailed study of SNR G296.5+10.0 and its CCO 1E~1207.4-5209 not only quantifies its contribution to the CR flux but also provides a natural laboratory for testing models of particle acceleration and propagation in low-luminosity environments, linking observational and theoretical results.

\begin{figure*}[!htb]
   \centering
   \subfloat[Radio image of the SNR G296.5+10.0]{\includegraphics[width=0.5\textwidth]{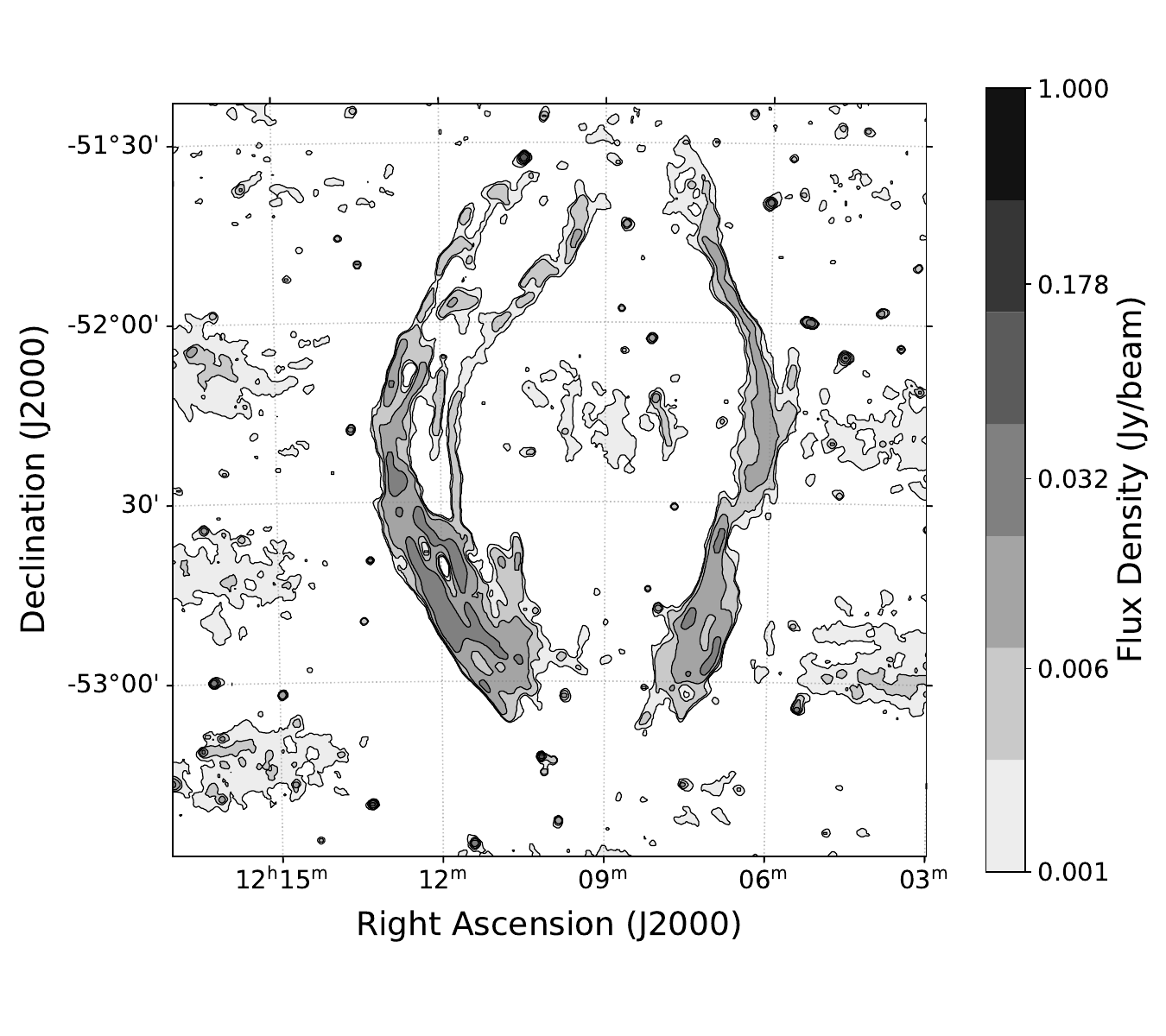}}
   \subfloat[ROSAT PSPC X-ray image of the SNR G296.5+10.0 region]{\includegraphics[width=0.5\textwidth]{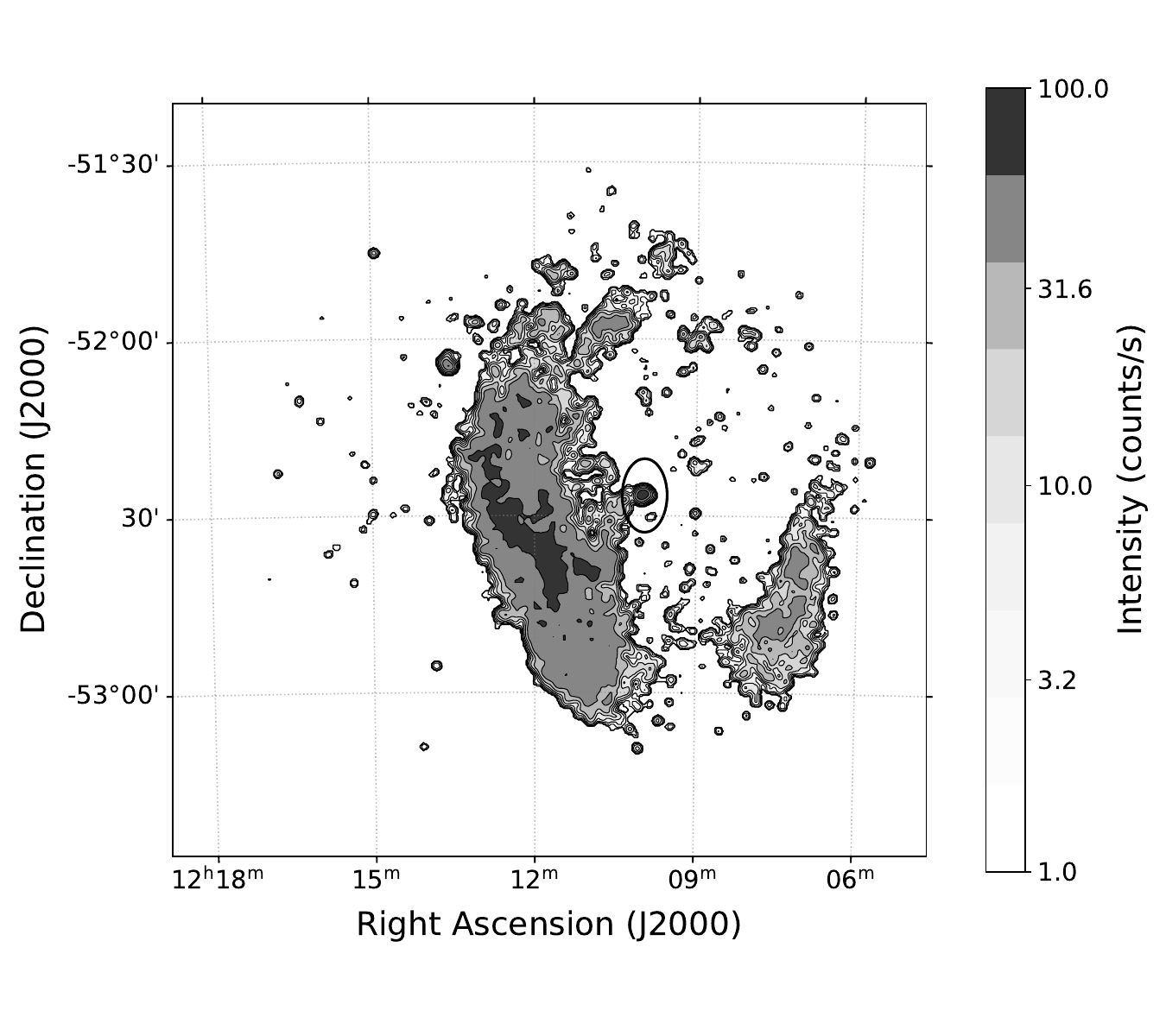}}\hfill
    \subfloat[TS map: 0.8 - 3\,GeV]{\includegraphics[width=0.5\textwidth]{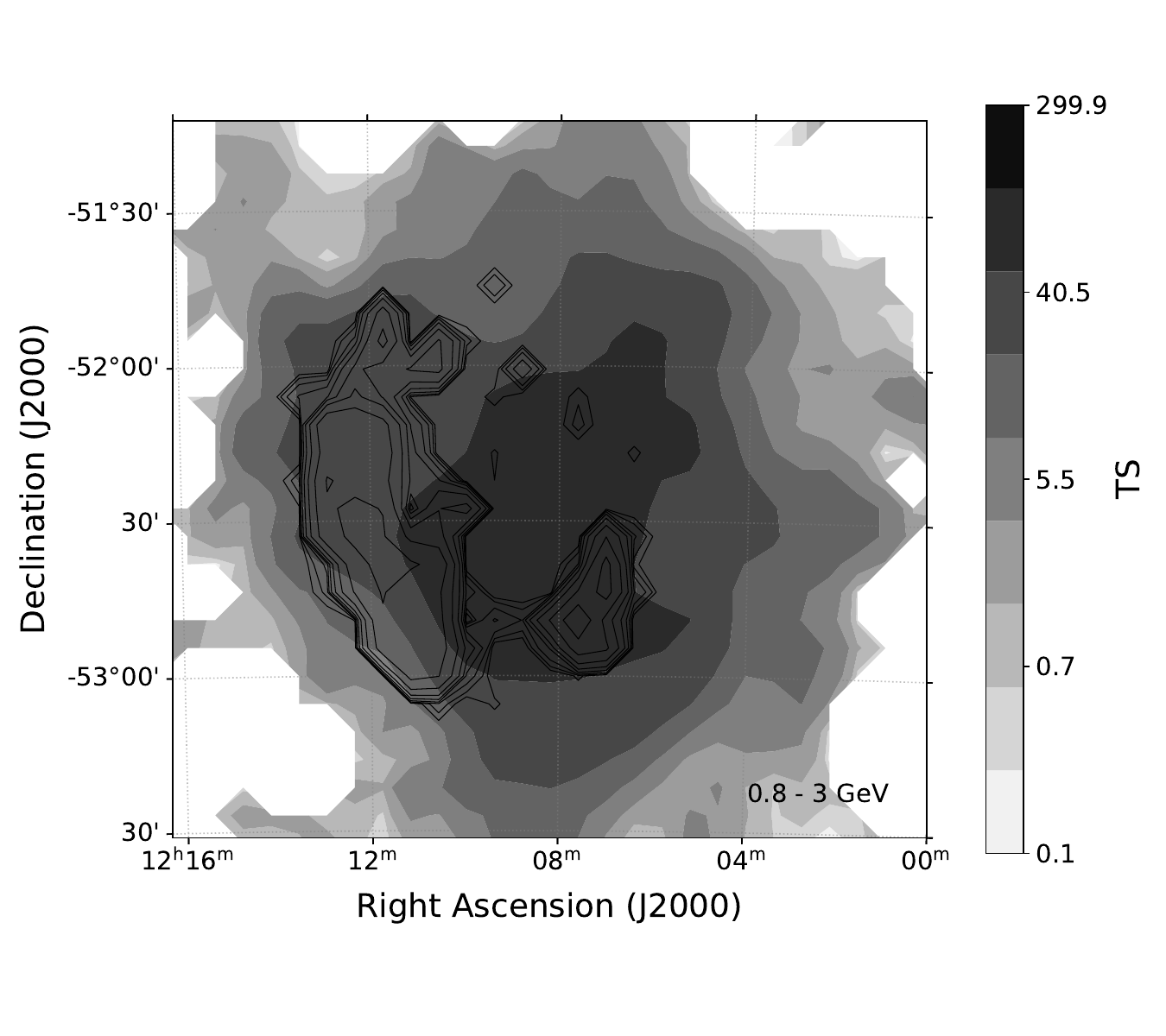}} 
    \subfloat[ TS map: 50 – 500\,GeV]{\includegraphics[width=0.5\textwidth]{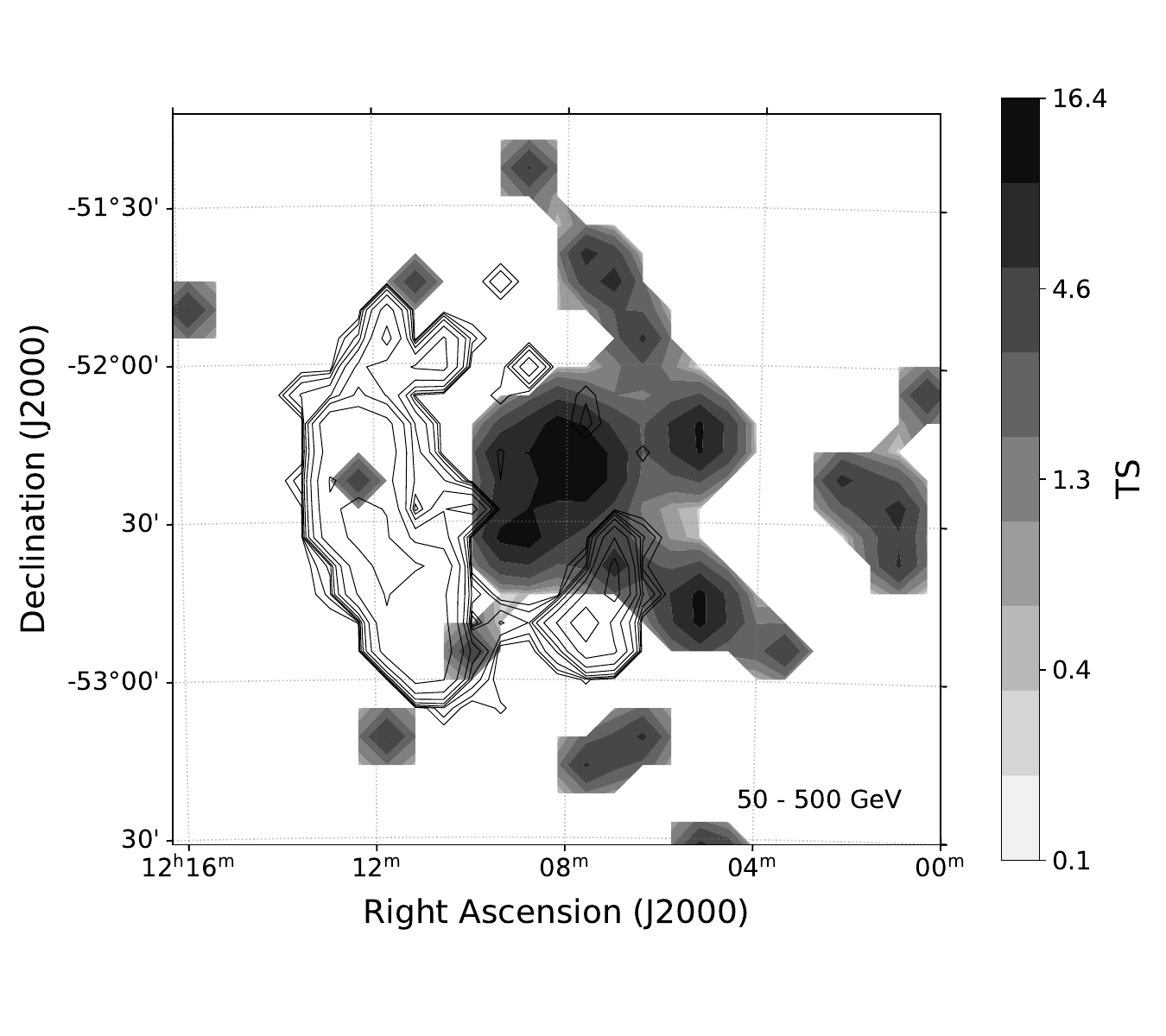}}\hfill
    \caption{(a) Radio continuum image of SNR G296.5+10.0 at 843 MHz, obtained from the SUMSS survey using the Molonglo Observatory Synthesis Telescope, showing its shell-like morphology. Contours indicate flux density levels from 0.001 to 1.000 Jy beam\(^{-1}\) (b) ROSAT PSPC X-ray image of SNR G296.5+10.0 in the 0.1–2.4 keV band (observation of 22 July 1993), processed with median and Gaussian filtering to enhance the signal-to-noise ratio. The black circle marks the CCO 1E~1207.4–5209 (c) and (d) TS maps of the gamma-ray emission toward SNR G296.5+10.0, computed from 14~yr of \textit{Fermi}-LAT Pass~8 data in the 0.8--3~GeV and 50--500~GeV bands, respectively, using the \texttt{TSMAP} method of the \texttt{fermipy} package \citep{2024MNRAS.528.2095E}. All colourbars in panels (a)–(d) are shown on a logarithmic scale.}
    \label{fits_x_radio_gamma}
\end{figure*}

This paper is structured as follows: Section~\ref{sec:CCO} describes the association between CCO 1E~1207.4-5209 and SNR G296.5+10.0, emphasizing their observational properties and implications for particle acceleration. Section~\ref{sec:simulations} details the simulation models and multimessenger analysis methodology, including the use of \texttt{GALPROP} for CR propagation. Section~\ref{sec:CTAO} presents the CTAO observational feasibility study, including sensitivity analyses and spectral modeling. Finally, in Sections~\ref{sec:results} and~\ref{sec:summary}, we discuss our results and their implications for understanding CR acceleration in SNRs and CCOs, with particular emphasis on the potential of next-generation gamma-ray observatories to probe these processes in greater detail.

\section{CCO 1E 1207.4-5209 and its host SNR G296.5 + 10.0}
\label{sec:CCO}

NS~1E~1207.4-5209 is a radio-quiet, fast-rotating object classified as a CCO \cite{2005PhRvL..95.1101G}. It was first identified near the core of the SNR~G296.5+10.0 (also known as PKS~1209-51/52) by the \textit{Einstein} satellite \cite{2000AJ....119..281G}. Located at the geometric centre of the remnant, it was observed in detail by \textit{XMM-Newton} in August~2002 \cite{2002ApJ...581.1280M}. The observation, carried out with the EPIC instrument and lasting approximately 36~hours, substantially improved the characterisation of its X-ray emission \cite{2004A&A...418..625D}. The CCO has a measured rotational period of $0.42413076\ \mathrm{s}$, with an upper limit on its period derivative of $2.224 \times 10^{-17}\ \mathrm{s\,s^{-1}}$ (according to the ATNF\footnote{ATNF Pulsar Catalogue: \url{https://www.atnf.csiro.au/research/pulsar/psrcat}}). Distance estimates for the system vary in the literature. For the SNR, Eppens et~al. \citep{2024MNRAS.528.2095E} report a distance of $1.4\ \mathrm{kpc}$, while the \href{http://snrcat.physics.umanitoba.ca}{SNRcat}\footnote{Supernova Remnant Catalogue: \url{http://snrcat.physics.umanitoba.ca}} catalogue lists a range of $1.3$--$3.9\ \mathrm{kpc}$. For the CCO, the ATNF lists $2.6\ \mathrm{kpc}$, but a value of $\sim 2\ \mathrm{kpc}$ is adopted in several other studies \cite{2000ApJ...540L..25Z,2015ApJ...812...61H,2011A&A...525A.106D,2005PhRvL..95.1101G,2002ApJ...578L.133H,2004A&A...418..625D,2000AJ....119..281G}. In this work, we adopt $2\ \mathrm{kpc}$, a widely used value that represents a reasonable average of the existing estimates. The X-ray light curve of 1E~1207.4-5209, as reported by De~Luca et~al. \cite{2004A&A...418..625D}, spans the energy range $0.2$--$4\ \mathrm{keV}$.

It is the second candidate for an Isolated Neutron Star (INS) discovered within a thermally emitting supernova remnant \cite{2011A&A...525A.106D}. The associated remnant, SNR G296.5+10.0, exhibits bright, structured emission in both X-ray and radio wavebands, concentrated particularly in its southern, eastern, and southwestern regions \cite{moranchel2017}. Its morphology has been interpreted as featuring tangentially aligned filaments of compressed cold gas \citep[see][]{roger1988} and a surrounding shell of neutral hydrogen ($\mathrm{H\,I}$) \cite{dubner}. A multiwavelength view of SNR~G296.5-+10.0 and the CCO 1E~1207.4-5209 is presented in Figure~\ref{fits_x_radio_gamma}. The composite image incorporates radio\footnote{Radio data retrieved from the NASA SkyView facility: \url{https://skyview.gsfc.nasa.gov}.}, X-ray\footnote{X-ray data from the ROSAT mission archives: \url{https://heasarc.gsfc.nasa.gov/docs/rosat/cdroms/rosatcdroms.html}.}, and gamma-ray observations \cite{2024MNRAS.528.2095E}. Panels (a) and (b) display the shell structure and central emission in radio and X-rays, respectively. The gamma-ray test statistic (TS) maps in panels (c) and (d) reveal energy-dependent morphological features that show spatial correlation with the X-ray contours \cite{2024MNRAS.528.2095E}. This comparative analysis provides evidence for the distribution of high-energy processes associated with the remnant and its central neutron star.

The pulsar 1E~1207.4-5209 holds particular significance as the first neutron star to display unambiguous, strong absorption lines in its spectrum. It has been proposed that a combination of youth and low magnetospheric activity creates favorable conditions for detecting atomic spectral features from elements with atomic number $Z > 1$ \cite{2002ApJ...581.1280M}. In older objects, such features would be obscured by an accreted hydrogen layer, while in younger, more energetic pulsars, they would be masked by dominant non-thermal processes. Subsequent modeling \cite{2002ApJ...578L.133H} constrained the atmospheric composition, indicating that the observed absorption features are naturally explained by the presence of oxygen or neon in a helium-like state within a magnetic field of order $\sim 10^{12}$~G. The associated magnetosphere is thus an environment rich in electron-positron pairs, with particles subject to acceleration via the pulsar's rotational spin-down energy.

The detection of absorption lines in 1E~1207.4-5209 enables a direct measurement of its magnetic field, unlike the case for most pulsars, whose field strengths are typically inferred indirectly from rotational energy loss models. This direct measurement offers a key constraint on the evolutionary state of this CCO. XMM-Newton observations reveal a cyclotron resonance feature at approximately 0.7~keV. Interpreting this as an electron cyclotron line implies a surface magnetic field of $\sim 10^{10}$~G \cite{2003Natur.423..725B,2004A&A...418..625D}. This relatively low field indicates that the NS is magnetically quiet, which has direct implications for age estimates. Using this measured field, the characteristic age of the CCO is three orders of magnitude larger than the age of the host supernova remnant, supporting models of magnetic field burial by fallback accretion \cite{2021ApJ...917...71Z, 2018ApJ...866..154G,2017JPhCS.932a2006D}. However, alternative estimates, derived from rotational parameters and from the interpretation of spectral lines as possibly arising from hadronic processes, suggest a significantly stronger magnetic field, exceeding $10^{10}$~G \cite{2010ApJ...709..436H,ankay2007,pavlov2005once,2004A&A...418..625D}. These conflicting results underscore the ongoing debate regarding the true magnetic field strength and configuration of 1E~1207.4-5209.

On the other hand, the physical association between CCO 1E~1207.4-5209 and SNR~G296.5+10.0 has been questioned. Studies such as that by de Luca et al. \cite{2011A&A...525A.106D} discuss the possibility of a chance alignment. The measured proper motion of the proposed optical counterpart is only $\sim7\ \mathrm{mas\ yr^{-1}}$, an order of magnitude lower than the $\sim70\ \mathrm{mas\ yr^{-1}}$ typical for CCOs. Furthermore, absolute astrometry comparing the optical source with \textit{Chandra} X-ray data reveals a significant positional offset, suggesting they are distinct objects. The high source density in the field also implies that the apparent proximity could be incidental Despite this, evidence for a buried magnetic field in 1E~1207.4-5209 exists. The detection of the first observed glitch in a CCO, studied by \cite{2018ApJ...866..154G}, supports this scenario. A glitch-a sudden increase in spin frequency can be explained by the gradual diffusion of an internal magnetic field toward the surface. This process generates stresses that interact with the internal neutron superfluid, triggering the event without altering the external dipole field. Such behavior aligns with models of a strong field buried by supernova fallback material. As suggested by Konstantinos et. al.~\cite{2022Symm...14..130G}, this submerged field may eventually resurge, potentially activating the star and enhancing particle emission into the magnetosphere.

In this work, we adopt the surface magnetic field derived from the cyclotron absorption lines detected by \textit{XMM-Newton}, interpreting them as electron transitions~\cite{2004A&A...418..625D}. This yields a field strength of $\sim 10^{10}$~G (accounting for a $25\%$ gravitational redshift) and a corresponding characteristic age of $\sim 10^{8}$~yr. This interpretation is the most consistent with current observations: the inferred low field is compatible with the measured spin-down, and the presence of four lines with an approximate 1:2:3:4 energy ratio strongly supports an electron cyclotron origin, which is difficult to reconcile with a proton interpretation. A proton cyclotron scenario would require a field of order $\sim 10^{14}$~G, inconsistent with the observed rotation parameters. Furthermore, the electron model naturally accommodates a suspended absorption layer located a few stellar radii above the surface, explaining both the lines' location and their phase-dependent variability while remaining consistent with a thin, localized absorbing region \cite{2004A&A...418..625D,2006ApJ...644..439L,2006ChA&A..30..263Y,2005MNRAS.356..359X,2002ApJ...574L..61S}. This choice allows us to model the system in its currently observed state as a magnetically quiet CCO with limited particle acceleration efficiency. Nevertheless, the possibility of a stronger, buried field remains relevant. If such a field gradually resurfaces, the pulsar could transition to a more active state, enhancing particle injection and potentially increasing the gamma-ray flux. While current data cannot fully test this hypothesis, future observations with facilities like the CTAO may provide the necessary constraints.

In the following, we model this system with two scenarios: a spin-down powered leptonic source associated with the CCO, and a quiescent hadronic source associated with the SNR shell. The corresponding \texttt{GALPROP} setup, including injection spectra and propagation parameters, is described in Section 3.

\section{Description of Cosmic-Ray Propagation Model}
\label{sec:simulations}

The latest version of \texttt{GALPROP} solves partial differential 
equations in three or four dimensions (accounting for spatial, energy, and time variables) using user-defined source distributions and boundary conditions. Its output includes intensity distributions for all major CR species in the interstellar medium (ISM)~\cite{galprop,1998ApJ...509..212S,2007ARNPS..57..285S}. The \texttt{GALPROP}~v57 simulations used in this work employed logarithmic energy grids spanning $1.0-10^9$~MeV per nucleon for CR and $1.0-10^{8}$~MeV for gamma rays. Let $N \equiv N(\vec{r}, p, t)$ denote the density of CR per unit momentum $p$ at position $\vec{r}$ in the Galaxy at time $t$. Under the assumption of isotropic diffusion, the CR transport equation can be expressed as \cite{galprop,1964ocr..book.....G}:
\begin{multline}
    \frac{\partial N}{\partial t} = Q(\vec r, p) 
    + \vec{\nabla} \cdot \left( D_{xx} \vec{\nabla} N - \vec{v} N \right) 
    + \frac{\partial}{\partial p} \left[ p^2 D_{pp} 
      \frac{\partial}{\partial p} \left( \frac{N}{p^2} \right) \right] \\
    - \frac{\partial}{\partial p} \left[ \dot{p} N 
      - \frac{p}{3} (\vec{\nabla} \cdot \vec{v}) N \right] 
    - \sum_{i=1}^{2} \frac{N}{\tau_i}.
    \label{eq:transport}
\end{multline}
where $Q(\vec r, p)$ is the source distribution of CR, including primary injection and secondary production. The $D_{xx}$ and $D_{pp}$ are the spatial and momentum diffusion coefficients, respectively, $\vec{v}$ is the advection velocity and $\tau_{\mathrm{r}}$ and $\tau_{\mathrm{f}}$ are the radioactive decay and fragmentation time scales. The advection velocity 
$\vec{v}(z)$ is assumed to increase linearly with the distance $z$ between the plane and the halo of the galaxy \cite{2007ARNPS..57..285S}. At low energies, up to $\sim 1$~GeV \cite{2016A&A...595A..33T}, both Galactic wind 
and energy losses play a significant role. A momentum diffusion coefficient $D_{pp}$ determines the re-acceleration. It is related to the spatial coefficient $D_{xx}$ via $D_{pp} \propto p^{2} v_{A}^{2}/D_{xx}$, where the Alfvén speed is denoted by $v_{A}$. By considering a Kolmogorov spectrum for interstellar turbulence, the diffusion coefficient is expressed as $D_{xx} = \beta D_{0,xx}(\rho/\rho_0)^{1/3}$, where $\beta$ is the particle velocity expressed in units of light speed, $D_{0,xx}$ is the 
normalization of the diffusion coefficient in the general ISM, $\rho$ is the particle magnetic rigidity, $\rho_0 = 3$~GV is the normalisation (reference) rigidity. We adopt a diffusion coefficient of $2.5 \times 10^{28}\,\rm cm^2\,s^{-1}$, consistent with standard Galactic propagation conditions and widely used in models \cite{2011ApJ...729..106T}. The Alfvén velocity is set to 28~km\,s$^{-1}$ to allow for moderate reacceleration, in agreement with the properties of the interstellar medium to reproduce the observed CR spectra \cite{2017A&A...597A.117D}. The transport equation is solved using the second-order Crank--Nicholson scheme.

\texttt{GALPROP} treats cosmic-ray nuclei on an isotope-by-isotope basis, solving the transport equation independently for each nuclear species characterized by its mass number \(A\) and charge \(Z\). The code incorporates a comprehensive nuclear reaction network, calculating primary, secondary, and tertiary CR species. All nuclei from $Z = 1$--$28$ are considered, and their evolution is tracked through the full nuclear reaction network to compute secondary and tertiary production. Light secondary-dominated species such as Li, Be, and B are set to zero at injection. In this work, particle spectra were extracted for individual isotopes by explicitly selecting the corresponding mass number \(A\), and no summation over isotopes was applied. The results correspond to the dominant stable isotope of each element, specifically, $^{4}$He, $^{12}$C, $^{14}$N, and $^{56}$Fe for the nuclear species. This 
comprehensive treatment enables accurate modeling of both stable and secondary CR species, including gamma-ray production from meson decay~\cite{2020ApJS..250...27B}.

Gamma-ray emissivities from gas-related processes, primarily $\pi^0$ decay and bremsstrahlung, are calculated using column density distributions of $\mathrm{H}_2$ (traced by CO) and HI for galactocentric rings, incorporating corrections for untraced gas 
\cite{2012ApJ...750....3A,2016ApJ...819...44A}. The ISM is predominantly composed of hydrogen and helium, with hydrogen existing in three main phases: HI (atomic, $\sim 60\%$), $\mathrm{H}_2$ (molecular, $\sim 25\%$), 
and HII (ionized, $\sim15\%$) \cite{2001RvMP...73.1031F}. The HII phase is typically low-density and diffuse, while $\mathrm{H}_2$ is concentrated in dense molecular clouds. Inverse Compton (IC) emissivities are calculated using anisotropic background photon distributions \cite{2000ApJ...528..357M}, while synchrotron emissivities are computed for both total and polarized 
components. Gamma-ray emission modeling includes hadronic $\pi^0$ decay following the formalism of Kamae et al. \cite{2006ApJ...647..692K}, bremsstrahlung from charged particles, and separate treatments for atomic and molecular hydrogen to account for their distinct contributions to the emission.

We model the Galactic CR propagation within a spatial domain extending $\pm18$~kpc in the $x$ and $y$ directions and $\pm6$~kpc in the $z$ direction. The grid is uniform in the $x$ and $y$ directions, with $\Delta x = \Delta y = 0.1$~kpc. In the $z$ direction, we use a non-uniform $\tan$ distribution, which gives a typical resolution of $\Delta z \sim 
0.03$~kpc near the Galactic plane and covers both the Galactic disk and halo \cite{galprop,2003PASJ...55..191N}. The source CCO~1E~1207.4-5209 is placed at coordinates RA$= 12^{\mathrm{h}}10^{\mathrm{m}}0.90^{\mathrm{s}}$, 
Dec$= -52^{\circ}26^{\prime}28.4^{\prime\prime}$ (ATNF), at a distance of $\sim2$~kpc. These coordinates are internally converted by \texttt{GALPROP} into Galactocentric Cartesian coordinates \((7.155,\allowbreak 1.762,0.344)\) assuming a solar position at 
\(R_{\odot}=8.5\,\mathrm{kpc}\).

For the injection spectrum, we adopt a differential power law in momentum, $dq(p)/dp \propto p^{-\alpha}$, and an  injection rate per unit momentum $q(p) \equiv d\dot{n}/dp$. In the test-particle limit of diffusive shock acceleration (DSA), the isotropic phase-space distribution is $f(p)\propto p^{-s}$ with $s=4$ for a strong shock. Since the differential number density per unit momentum is $n(p)=4\pi p^{2} f(p)$, the source term expressed per unit momentum has an index $\alpha = s-2 = 2$. We therefore take $\alpha=2.0$ as our fiducial value and explore $\alpha=1.8$ and $\alpha=2.2$ as reasonable bounds. This range is chosen to match gamma-ray observations under the assumption of a single, localized source 
associated with CCO~1E~1207.4-5209 and SNR~G296.5+10.0. Slightly harder spectra ($\alpha<2$) can arise when shock modification is minimal or when magnetic-field amplification and feedback are weaker. Because we model a single young, isolated source rather than an ensemble, exploring modest hardening is appropriate to assess the maximal contribution to the local CR and gamma-ray fluxes. Moreover, a harder index improves the fit to the highest-energy gamma-ray data while remaining within a physically motivated range supported by DSA theory and supernova remnant observations~\cite{2012JCAP...07..038C,2021ApJ...922....1D,2013MNRAS.434.2202A}.

Energy losses are calculated from the perspective of an observer at the solar system. For the interstellar radiation field (ISRF) we employ the R12 model, a three-dimensional representation of the Milky Way based on smooth distributions of stellar populations and interstellar dust~\cite[see][]{2012A&A...545A..39R,porter2017}. The Galactic magnetic field is described by a basic exponential model with cylindrical symmetry~\cite{sun2008}:
\begin{equation}
    B(r,z) = B_{0} \, e^{(R_{\odot}-r)/R_{B}} \, e^{-\left|z \right|/z_{B}},
\end{equation}
where $r$ is the Galactocentric radius, $R_{\odot}=8.5$~kpc is the solar radius, $B_{0}=5\,\mu\mathrm{G}$, and $R_{B}=6$~kpc and $z_{B}=2$~kpc are the radial and vertical scale lengths~\cite{2000ApJ...537..763S,1998ApJ...509..212S}. This model accounts only for the random  magnetic-field distribution, which governs the diffusive transport of cosmic rays~\cite{2017JCAP...02..015E,2020ApJ...892....6L}. Throughout the simulation we solve for the CR density per unit momentum $N(\mathbf{r},p,t)$ with units $\mathrm{cm^{-3}\,(GeV/c)^{-1}}$, using a source term $Q(\mathbf{r},p)=n(\mathbf{x})\,q(p)$ with units
$\mathrm{cm^{-3}\,s^{-1}}$ \allowbreak $\mathrm{(GeV/c)}^{-1}$.\footnote{When expressed in  kinetic energy $E$, the corresponding units are $\mathrm{cm^{-3}\,s^{-1}\,GeV^{-1}}$.} All parameters described above are applied consistently in both models detailed below.\footnote{The data supporting this study are available from the corresponding author upon reasonable request.}

\subsection{Spin-down model}

In this scenario, the rotational energy loss of CCO~1E~1207.4-5209 through dipole radiation accelerates electrons and positrons, which are continuously injected into the surrounding medium. The spin-down model includes only electrons and positrons, assuming that CCOs predominantly accelerate leptons and do not contribute significantly to nuclear CR components. The total injection power is normalized to the spin-down luminosity:
\begin{equation}
L(t) = \eta L_{0} \left(1 + \frac{t}{\tau_0}\right)^{-2},
\label{power}
\end{equation}
where \(L_0 = 1.0 \times 10^{33}~\mathrm{erg\,s^{-1}}\) is the initial rotational power of the source \cite{2004A&A...418..625D,2000ApJ...540L..25Z}, 
\(\eta = 1\) is the particle acceleration efficiency (assumed 100\%), and \(\tau_0 = 3.63 \times 10^{5}~\mathrm{yr}\) is the characteristic age derived from the CCO's magnetic field. The age of the source is taken as \(t = 3.02\times10^{8}\)~yr (according to the ATNF). The injection spectrum itself follows a power law with a smooth break, and the rate of CR production is governed by the pulsar's spin-down evolution~\cite{coelho2022updated,malyshev2009pulsars}.

\subsection{Quiescent model}

The quiescent model considers hadronic emission from the SNR~G296.5+10.0 shell, in which accelerated protons lose energy through hadronic (p--p) interactions, producing neutral pions that decay into gamma rays. This is the primary mechanism for high-energy emission from such remnants. VHE emission can also arise from interactions involving protons, electrons, ambient matter, and radiation fields. In the quiescent scenario, electrons 
and positrons are injected with relative abundances of \(1.06\times10^3\) and 1.06, respectively. Hydrogen and helium dominate the nuclear component, while heavier nuclei up to nickel are included with abundances following \texttt{GALPROP} standard normalization at a reference kinetic energy.

To assess the time evolution of the CR luminosity, we simulate different source ages $t_{\mathrm{age}}$, representing the time elapsed since the onset of CR injection. We consider three values: $t_{\mathrm{age}} = 5\times10^{4}$, $10^{5}$, and $3\times10^{6}$~yr. For each value, \texttt{GALPROP} is run in steady-state mode with the source luminosity normalized to the corresponding age, effectively bracketing plausible evolutionary stages of the system. These timescales were chosen to probe both the estimated age of the SNR ($\sim10^{4}$~yr~\cite{2002ApJ...569L..95P}) and its long-term evolution toward a steady-state configuration. This allows us to evaluate the expected CR and gamma-ray flux from SNR~G296.5+10.0 at Earth at different evolutionary stages. While the younger ages test the time-limited propagation of particles from a recent injection event, the longest simulation time approximates an equilibrium scenario and sets an upper bound on the potential contribution of this SNR to the local CR flux.

Extragalactic propagation models suggest that the relationship between the total flux of GeV--TeV gamma rays emitted by a source and the propagation of CR can establish an upper limit on the overall CR luminosity~\cite{Supanitsky_2013, Anjos_2014, 2025ApJ...994...31S}. In cases where only an upper limit on the integrated gamma-ray flux is available, such as G296.5+10.0, this limit 
constrains the total primary and secondary emission from the source. We assume that the measured gamma-ray upper limit reflects the maximum allowed secondary photon flux produced by CR propagating from the source to Earth, enabling us to derive a corresponding upper bound on the source's CR luminosity. Under the assumption of isotropic CR emission, the gamma-ray intensity at Earth can be expressed as:
\begin{equation}
I_{\gamma}(E_{\gamma}) \sim \frac{W_{\mathrm{CR}}}{4\pi d^{2} 
\langle E_{0} \rangle} K_{\gamma} P_{\gamma}(E_{\gamma}),
\end{equation}
where $W_{\mathrm{CR}}$ is the total energy injected into CR nuclei by the source, $d$ is the distance to the source (here, 2~kpc), $\langle E_0 \rangle$ is the average energy per CR particle, $K_{\gamma}$ is the gamma-ray production yield per CR interaction and $P_{\gamma}(E_{\gamma})$ is the normalized energy distribution of secondary gamma rays observed at Earth~\cite{2021JCAP10023D,coelho2022updated,Sasse:2022pqn,
Mocellin:2022occ}. We use the 99.5\% confidence level upper limit on the integral gamma-ray flux in the range $1\,\mathrm{GeV} < E < 1\,\mathrm{TeV}$, reported by \textit{Fermi}-LAT as $(1.19 \pm 0.11) \times 10^{-9}\ \mathrm{cm^{-2}\,s^{-1}}$~\cite{2021ApJ...910...78Z}. Assuming that all observed gamma-ray emission in this energy band is of secondary origin from CR interactions, this flux serves as an upper bound for the CR luminosity from the association. By comparing modeled gamma-ray spectra with this observational upper limit, we constrain the total CR energy budget that could have been injected by the CCO$+$SNR association and propagated to Earth within the corresponding timeframe.

\section{Observational Performance of the CTAO}\label{sec:CTAO}

The absence of pulsed gamma-ray detection from the pulsar 1E 1207.4-5209 suggests that gamma-ray emission in the region may be suppressed due to absorption by the nearby SNR G296.5+10.0. A recent high-energy study by Eppens et al. \cite{2024MNRAS.528.2095E}, based on 14 years of Fermi-LAT data, indicates that the gamma-ray emission from G296.5+10.0 is predominantly produced by hadronic interactions, with no significant energy-dependent morphological variations across different gamma-ray bands. The study also reveals a hard, unattenuated gamma-ray spectrum extending up to 500 GeV without an explicit cutoff, suggesting efficient CR acceleration within the remnant and supporting the idea that interactions with the surrounding medium, including dense clouds, shape the observed high-energy emissions. However, with the advent of advanced instruments such as CTAO, the prospects for detecting gamma rays from this pulsar are improving. CTAO’s enhanced sensitivity will enable the detection of weaker and less prominent sources, which is crucial for revealing subtle emission from this region.

\begin{table}[!ht]
    \begin{minipage}{0.5\textwidth}
    \centering
     \caption{Celestial coordinates in the ICRS using Right Ascension (R.A.) and Declination (Dec.) of the counterparts used in the simultaneous likelihood fit, as well as the angular separation (Sep.) from the center ($\mathrm{R.A.} = 182.13^\circ,  
 \mathrm{Dec.} = -52.73^\circ$) \cite{Acero_2016} of the SNR G296.5+10.0 region.  
    }\label{tab:Counterparts}
    \begin{tabular}{llll}
    \toprule
    \textbf{Counterparts} & \textbf{R.A.} & \textbf{Dec.} & \textbf{Sep.} \\
    \midrule
    Fermi-LAT [3FHL] (2017) \cite{Ajello_2017} & 182.13$^\circ$ & -52.73$^\circ$ & 0.003$^\circ$ \\ 
    % 3FGL J1209.1-5224 \cite{Acero_2015} & 182.28$^\circ$ & -52.40$^\circ$ & 0.336$^\circ$ \\
    Fermi-LAT [4FGL] (2022) \cite{2020ApJS..247...33A, Abdollahi_2022} & 182.13$^\circ$ & -52.73$^\circ$ & 0.003$^\circ$ \\
    % ARAYA (2013) \cite{2013MNRAS.434.2202A} & 182.25$^\circ$ & -52.45$^\circ$  &  0.600$^\circ$ \\
     \bottomrule
    \end{tabular}
         \end{minipage}
\end{table}

In this section, we assess the CTAO's capacity to detect gamma-ray emissions from the SNR G296.5+10.0 surroundings, applying methodologies from \cite{Costa2024} and related works \cite{Sousa_2025, 2025BrJPh..55...60S}. The \texttt{Gammapy}\footnote{\url{https://gammapy.org/}} program was utilized to perform these analyses \cite{Deil_2017, Donath_2023}.

We modeled the source as centered on the position of SNR G296.5+10.0 ($\mathrm{R.A.} = 182.13^\circ, \mathrm{Dec.} = -52.73^\circ$) \footnote{In the International Celestial Reference System (ICRS), using Right Ascension (R.A.) and Declination (Dec.).} \cite{Acero_2016}, covering the multi-GeV to multi-TeV range. To obtain the spectral parameters for our CTAO forecasts, we performed a simultaneous maximum-likelihood fit to data from the $3^{\rm{rd}}$ Hard Fermi-LAT Source Catalog (3FHL) \cite{Ajello_2017} and the $4^{\rm{th}}$ Fermi-LAT Source Catalog (4FGL) \cite{Abdollahi_2022, 2020ApJS..247...33A}. The resulting normalization $\Phi_0$ and photon index $\alpha$ (listed in Table~\ref{tab:skymodel}) are fully consistent, within uncertainties, with the hard, unattenuated spectrum reported by Eppens et al. up to $\sim500$ GeV~\cite{2024MNRAS.528.2095E}. Although the \texttt{Fermi}-LAT data are well described by a simple power law with no statistically significant cutoff within the LAT energy range, a high-energy cutoff is physically expected for a finite-aged accelerator. To obtain a realistic extrapolation into the CTAO band, we adopt an Exponential Cutoff Power-Law (ECPL) model:
\begin{equation}
\Phi(E) = \Phi_0\ (E/E_{0})^{-\alpha} e^{-E/E_{\rm{cut}}},
\label{eq:ecpl}
\end{equation}
where $E_0 = 0.1~\rm{TeV}$ is the reference energy, and $E_{\rm{cut}}$ is the cutoff energy. We set a maximum cutoff energy of $20\ \mathrm{TeV}$, consistent with theoretical expectations from diffusive shock acceleration and TeV observations of similar sources, to obtain a physically reasonable extrapolation into the CTAO band~\cite{2019scta.book.....C}. To assess the expected performance of CTAO in detecting the modeled source, we employed the 1D ON/OFF observation technique, as described in \cite{Piano_2021}. The ON region (centered on the source) was defined as a circular area with a radius of $0.3^\circ$, while the OFF region (background) was determined using the Instrument Response Functions (IRFs) provided by the CTA Consortium and CTAO (version prod5 v0.1 \cite{CTAOIRFS}), with a scaling factor of $\alpha = 0.2$. We assumed parallel telescope pointing and implemented a wobbling mode with a $0.5^\circ$ offset around the source position. A $0.3^\circ$ ON region is appropriate for CTAO \footnote{\url{https://www.cta-observatory.org/science/cta-performance/}}, as it encompasses the expected point spread function across energies while optimizing background rejection \cite{2007A&A...466.1219B}. 

\begin{table}[!ht]
    \begin{minipage}{0.48\textwidth}
    \centering
     \caption{The position in the ICRS using Right Ascension (R.A.) and Declination (Dec.), radius and spectral model for the SNR G296.5+10.0 region. The modeled spectral parameters are derived from the likelihood fit, whereas the simulated parameters are based on the CTAO observation.}\label{tab:skymodel}
        \begin{tabular}{lll}
        \toprule
        position (R.A., Dec.)  \\ 
         (deg) & \multicolumn{2}{c}{($182.13, -52.73$) \cite{Acero_2016}} \\ 
          radius (deg.) & \multicolumn{2}{c}{ 0.76 \cite{Acero_2016}} \\ 
          \midrule
         fit & likelihood & CTAO South (50h)\\
            \cmidrule(l){2-3} 
        amplitude $\Phi_0 \times 10^{-10}$           \\ 
        ($\mathrm{cm^{-2}\,s^{-1}\,TeV^{-1}}$) &
        $1.43 \pm 0.21$ & $1.41 \pm 0.07$ \\
        index $\alpha$  & 
        $1.908 \pm 0.044$ & $1.904 \pm 0.024$ \\
        cutoff energy $E_{\mathrm{cut}}$\\
        ($\mathrm{TeV}$) & 
        $19.99 \pm 3.50$  & $18.51 \pm 2.56$ \\
         \bottomrule
        \end{tabular}
    % \tablerefs{$^a$In the International Celestial Reference System (ICRS) using Right Ascension (R.A.) and Declination (Dec.).}
     \end{minipage}
\end{table}

We set a minimum requirement of 10 expected signal counts and a significance threshold of $5\sigma$ per bin \cite{CTAOIRFS}. Based on this configuration, we generated energy-dependent differential flux sensitivity curves for both full-array and subarray IRF configurations, assuming a 50 hr observation time. The integral sensitivity was computed over the energy range of 100 GeV to 32 TeV. Table~\ref{tab:Counterparts} presents the celestial coordinates of the counterparts included in the simultaneous likelihood fit, along with their angular separations from the center of the region. In addition, Table~\ref{tab:skymodel} summarizes the parameters of the spectral model for the SNR G296.5+10.0 region, as derived from the simultaneous likelihood fit. The table includes key characteristics of the modeled source, such as its celestial coordinates and spatial extent.

\begin{figure*}[!htb]
   \centering
   \subfloat[Spectral indices for quiescent model]{\includegraphics[width=0.51\textwidth]{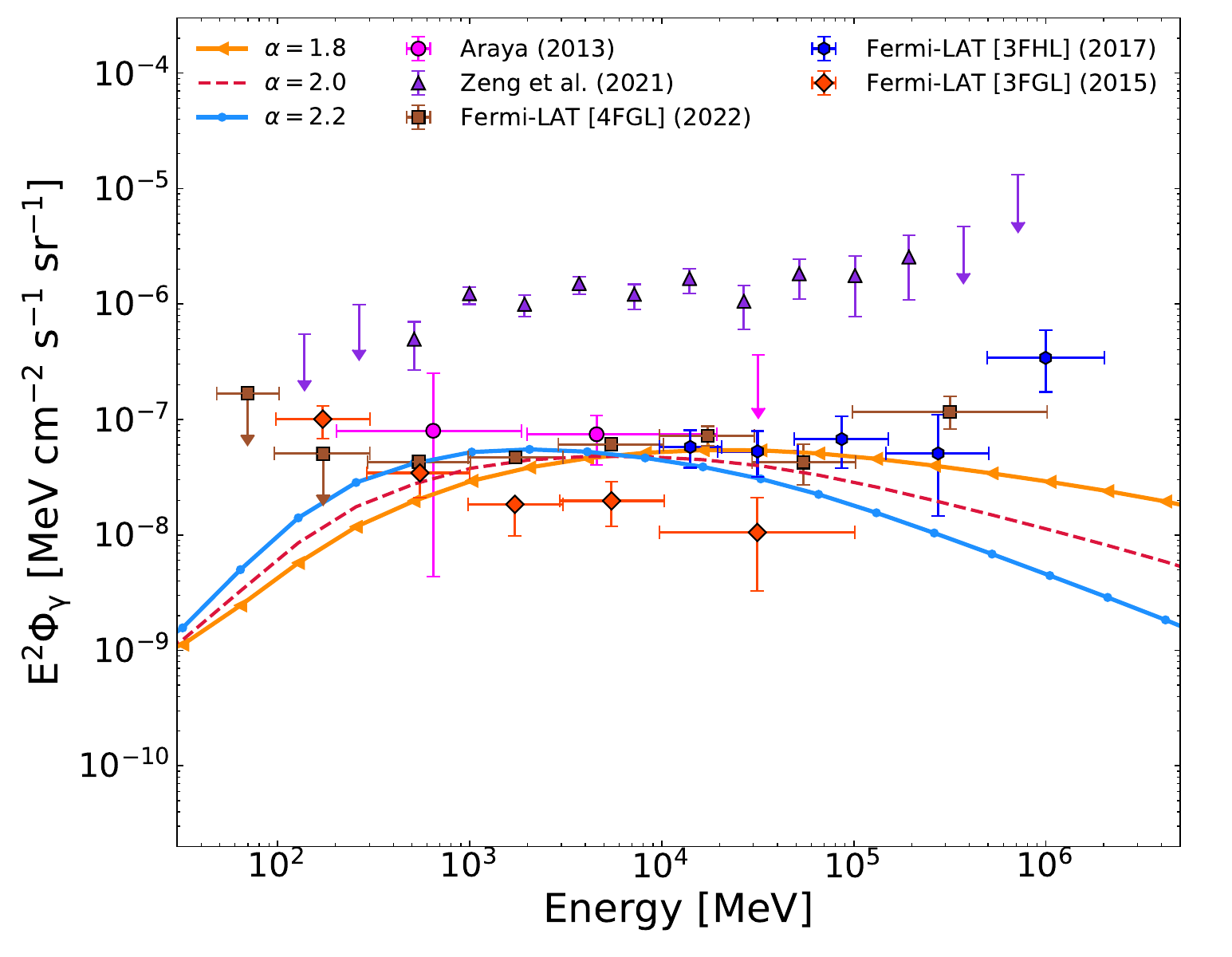}} 
   \subfloat[Spin model for $\alpha = 1.8$]{\includegraphics[width=0.51\textwidth]{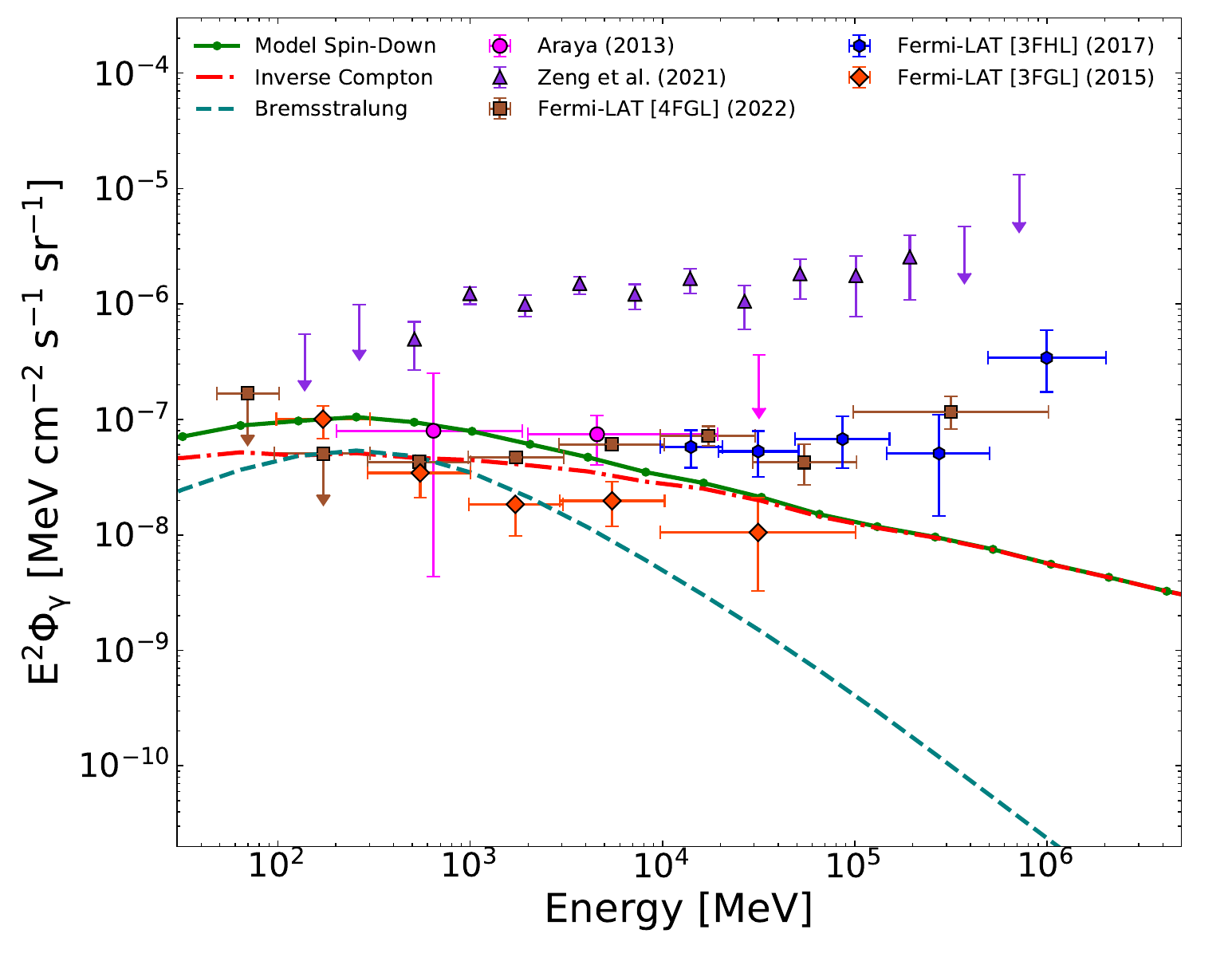}}\hfill   
%   \subfloat[Quiescent model for $\alpha = 1.8$]%{\includegraphics[width=0.41\textwidth]{quiescent_total_1.8.pdf}}
 %  \subfloat[Quiescent model for different times and $\alpha = 1.8$]{\includegraphics[width=0.41\textwidth]{Times_1.8.pdf}}\hfill
%   \subfloat[Spin and Quiescent models for $\alpha = 1.8$]{\includegraphics[width=0.41\textwidth]{Quiescent_spin_1.8.pdf}}
   \caption{Simulated gamma-ray spectral energy distributions for the CCO 1E~1207.4-5209 and its host SNR G296.5+10.0, compared with observational data from Fermi-LAT (3FGL, 3FHL, 4FGL), Araya (2013), and Zeng et al. (2021) \cite{2013MNRAS.434.2202A,Ajello_2017,Abdollahi_2022,2020ApJS..247...33A,Acero_2015,2021ApJ...910...78Z}.(a) Quiescent model for different spectral indices ($\alpha = 1.8$, $2.0$, $2.2$), evaluated at a fixed elapsed propagation time of $5 \times 10^4$ yr. (b) Spin-down model assuming steady-state leptonic injection from the CCO, showing the inverse Compton and bremsstrahlung components.}
    \label{gamma_models_1}
\end{figure*}

\begin{figure*}[!htb]
   \centering
%   \subfloat[Spectral indices for quiescent model]{\includegraphics[width=0.41\textwidth]{index_quiescent.pdf}} 
%   \subfloat[Spin model for $\alpha = 1.8$]{\includegraphics[width=0.41\textwidth]{Spin.pdf}}\hfill   
%   \subfloat[Quiescent model for $\alpha = 1.8$]{\includegraphics[width=0.41\textwidth]{quiescent_total_1.8.pdf}}
   \subfloat[Quiescent model for different times and $\alpha = 1.8$]{\includegraphics[width=0.51\textwidth]{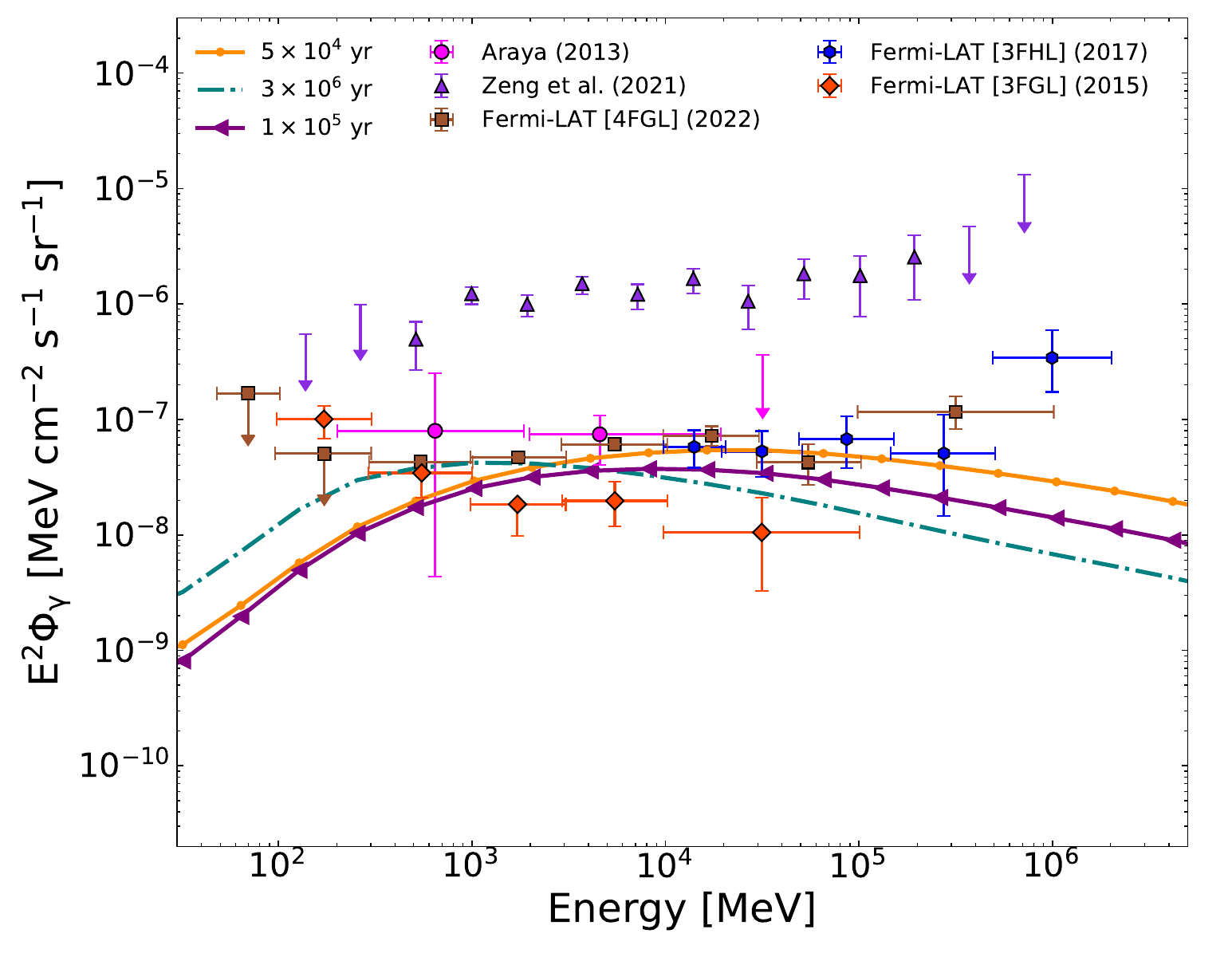}}
   \subfloat[Quiescent model for $\alpha = 1.8$]{\includegraphics[width=0.51\textwidth]{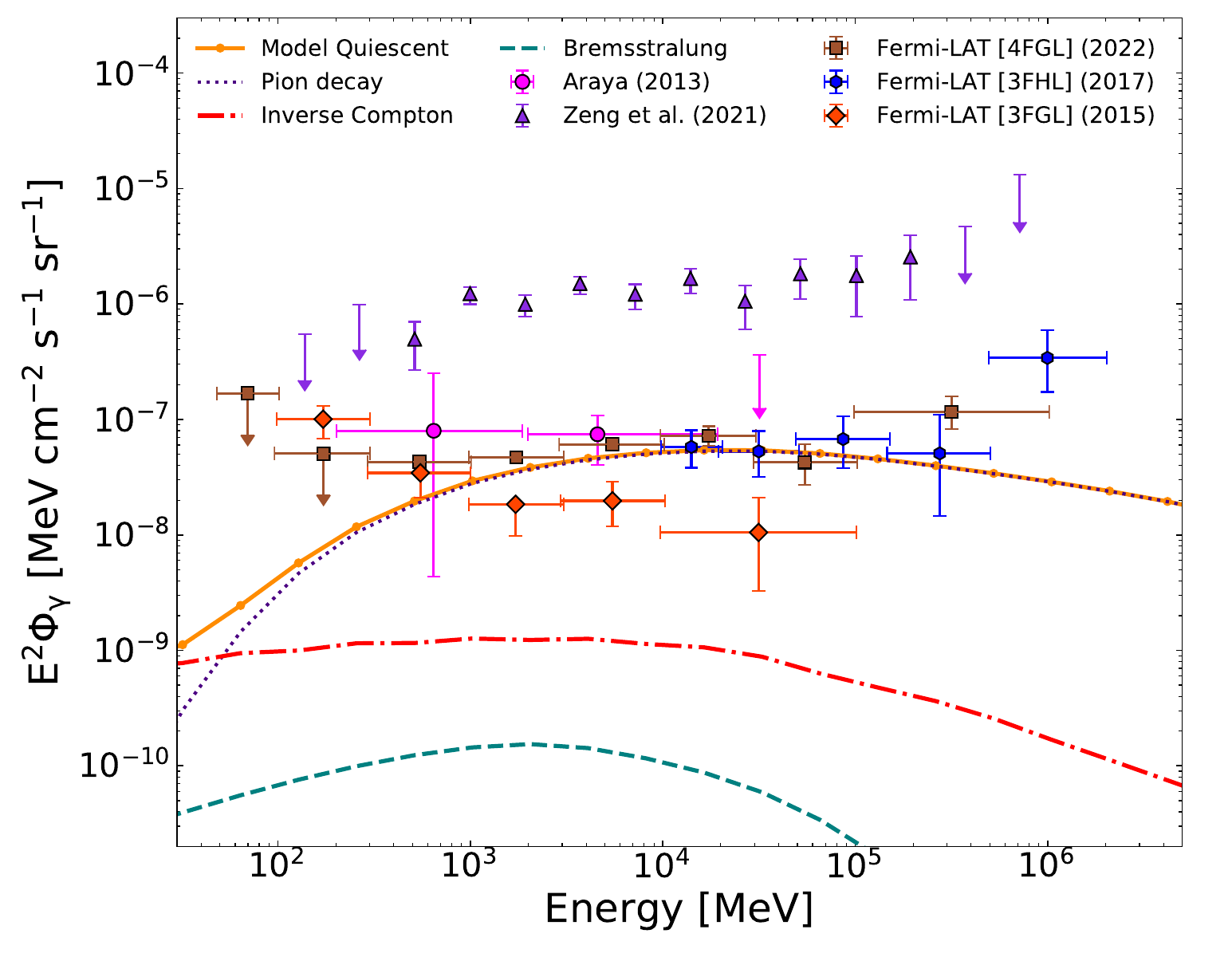}}
   \caption{Simulated gamma-ray spectral energy distributions for the CCO 1E~1207.4-5209 and its host SNR G296.5+10.0, compared with observational data from Fermi-LAT (3FGL, 3FHL, 4FGL), Araya (2013), and Zeng et al. (2021) \cite{2013MNRAS.434.2202A,Ajello_2017,Abdollahi_2022,2020ApJS..247...33A,Acero_2015,2021ApJ...910...78Z}.(a)  Time evolution of the quiescent model at $\alpha = 1.8$, for elapsed propagation times of $5 \times 10^4$, $10^5$, and $3 \times 10^6$ yr. (b) Quiescent model for elapsed propagation time of $5 \times 10^4$ yr and showing leptonic and hadronic components. }
    \label{gamma_models_2}
\end{figure*}

\begin{figure*}[!htb]
   \centering
   \includegraphics[width=0.55\textwidth]{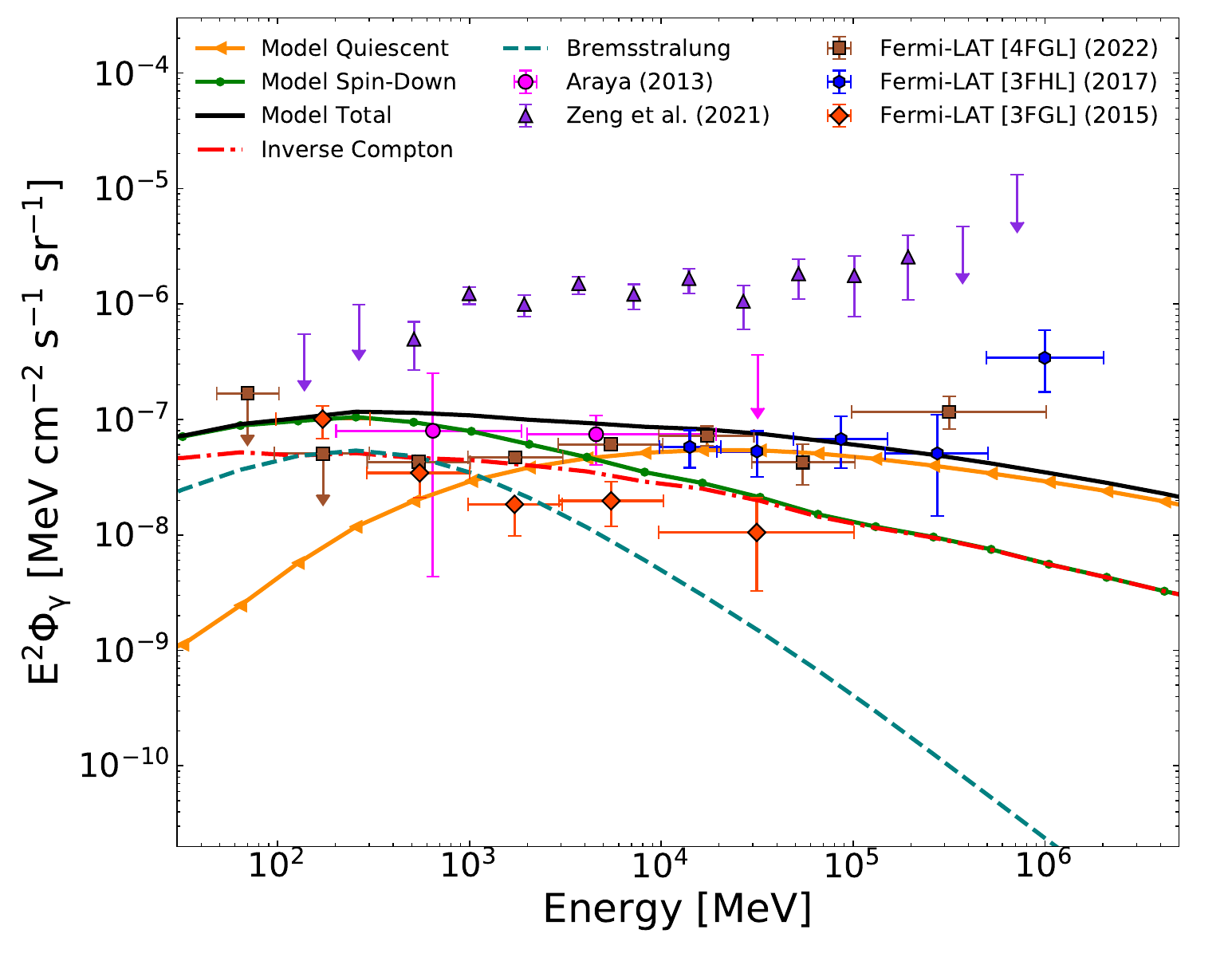}
   \caption{Simulated gamma-ray spectral energy distributions for the CCO 1E~1207.4-5209 and its host SNR G296.5+10.0, compared with observational data from Fermi-LAT (3FGL, 3FHL, 4FGL), Araya (2013), and Zeng et al. (2021) \cite{2013MNRAS.434.2202A,Ajello_2017,Abdollahi_2022,2020ApJS..247...33A,Acero_2015,2021ApJ...910...78Z}. Combined emission model including both the time-evolving hadronic (quiescent) and steady-state leptonic (spin-down) components. The CCO contributes primarily at lower energies via IC scattering, while the SNR dominates the high-energy emission through hadronic interactions.}
   \label{gamma_models_3}
\end{figure*}

\begin{figure*}[!ht]
   \centering
   \subfloat[$\alpha = 1.8$]{\includegraphics[angle=0,width=0.51\textwidth]{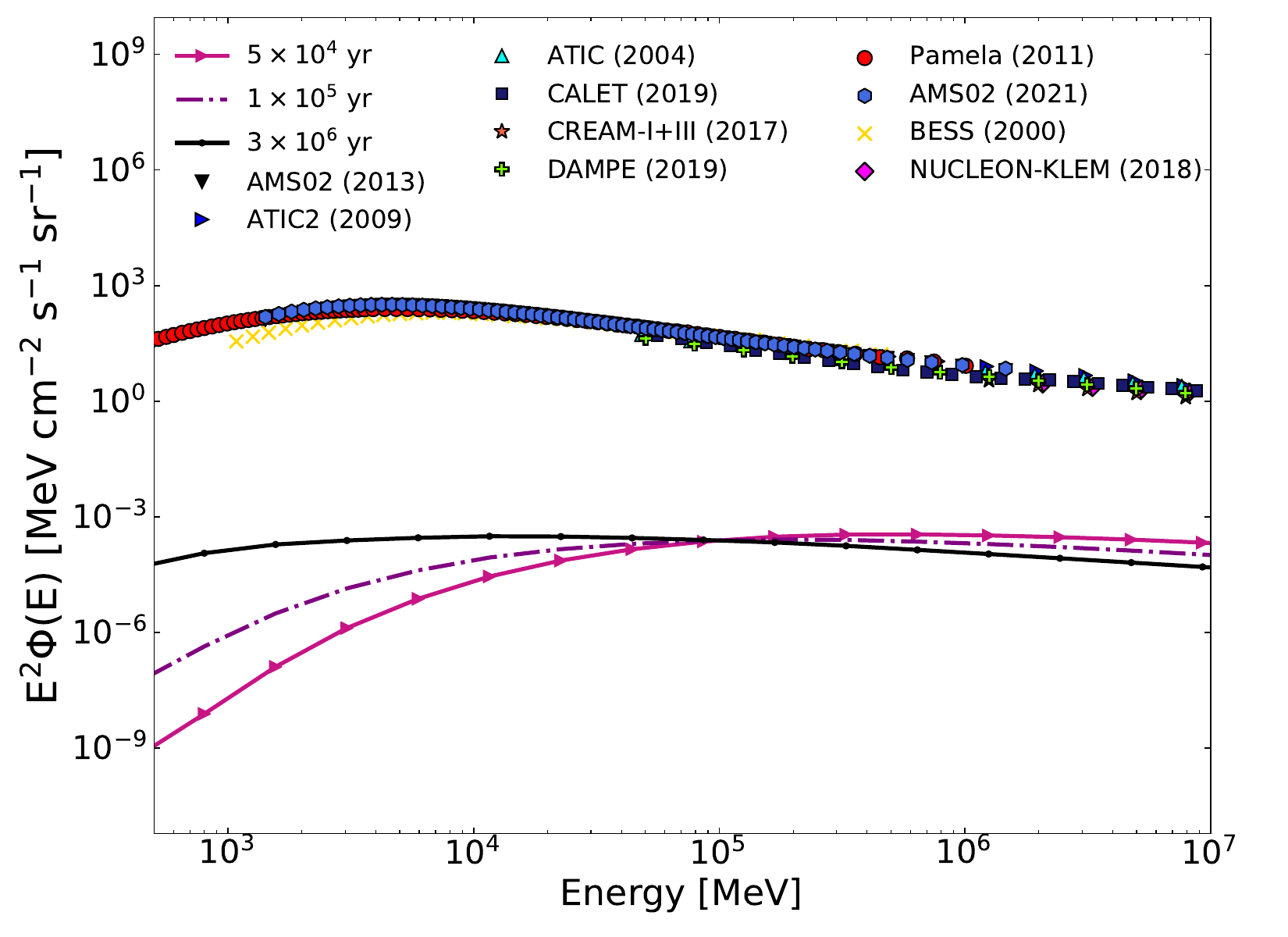}}
   \subfloat[$\alpha = 2.0$]{\includegraphics[angle=0,width=0.51\textwidth]{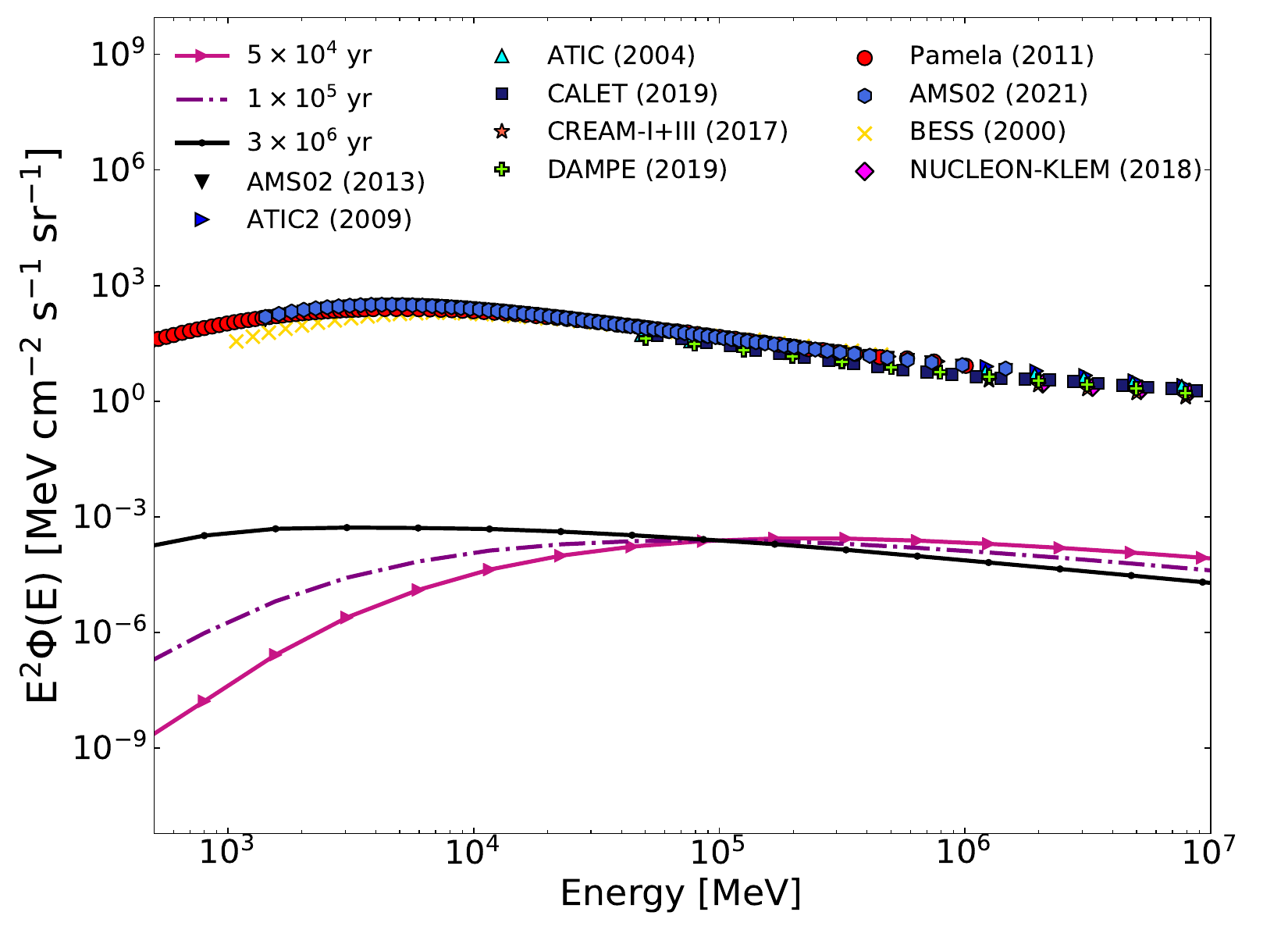}}\\
   \subfloat[$\alpha = 2.2$]{\includegraphics[angle=0,width=0.51\textwidth]{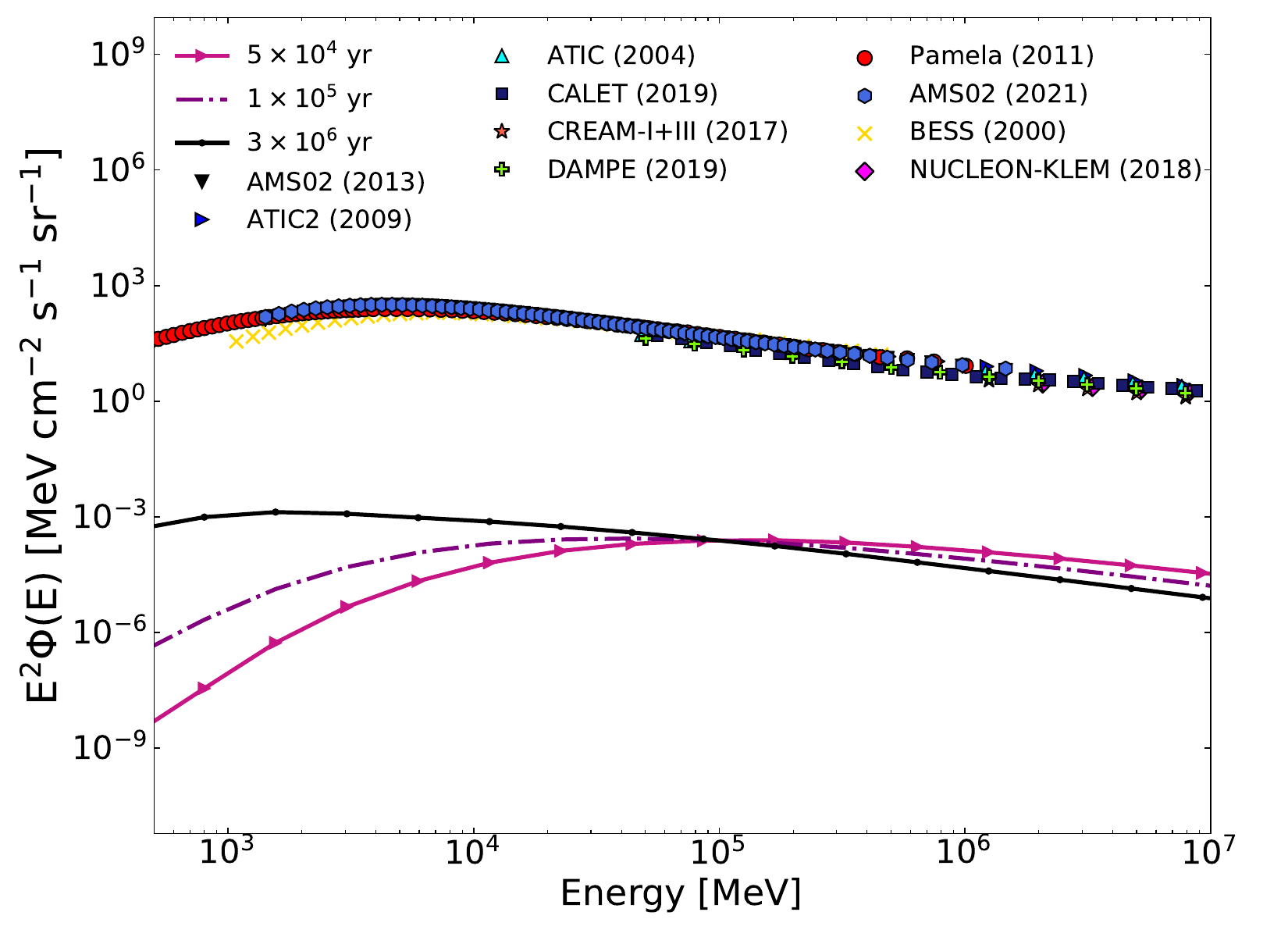}}
   \caption{Energy spectra of protons predicted by the quiescent emission model for different times and spectral indices $\alpha = 1.8$, $2.0$, and $2.2$ (panels a–c, respectively), multiplied by $E^2$. A modulation potential of 0.30~GV is applied. These spectra illustrate how variations in the spectral index influence the CR proton distribution from SNR G296.5+10.0. The data have been extracted from~\cite{aguilar2021,sanuki2000proton,2015PhRvL.114q1103A,dampe2019,yoon2017proton,atkin2018,2009BRASP..73..564P}.}
    \label{spc_proton}
\end{figure*}

\section{Discussions}\label{sec:results}

We analyze the SEDs of five nuclear species (proton, helium, carbon, nitrogen, and iron), along with electrons and positrons, within two primary cosmic-ray injection scenarios: the quiescent model and the spin-down model (detailed in Section~\ref{sec:simulations}). The quiescent model assumes a mixed composition of CR nuclei and positrons and electrons, representative of standard SNR injection processes. In contrast, the spin-down model includes only electrons and positrons, based on the assumption that CCOs predominantly accelerate leptons and do not contribute significantly to nuclear CR components. Both models incorporate a two-dimensional distribution of interstellar gas to account for spatially dependent gamma-ray production via hadronic and leptonic interactions \cite{johannesson2018, porter2017}.

Figures~\ref{gamma_models_1}, \ref{gamma_models_2} and \ref{gamma_models_3} present the simulated gamma-ray SEDs for the SNR~G296.5+10.0 and its associated CCO~1E~1207.4-5209, evaluated for spectral indices $\alpha = 1.8$, $2.0$, and $2.2$ and for different elapsed propagation times $5 \times 10^4$, $10^5$, and $3 \times 10^6$ yr. The panels display gamma-ray fluxes from two cosmic-ray injection scenarios: the quiescent model, which assumes hadronic emission from the SNR shell evaluated 
at different source ages $t_{\mathrm{age}}$, and the spin-down model, which assumes steady-state leptonic injection from the CCO. The total flux is the sum of both contributions. Additionally, individual components from Inverse Compton (IC) scattering and bremsstrahlung, exclusive to the spin-down model, are shown. These results are compared with Fermi-LAT data (3FGL, 3FHL, 4FGL) and previous measurements reported in \cite{2013MNRAS.434.2202A,Ajello_2017,Abdollahi_2022,2020ApJS..247...33A,Acero_2015}. With the exception of Figure~\ref{gamma_models_2}-(b), all panels showing the quiescent model are calculated for a elapsed propagation time of $5 \times 10^4$ yr; panel (b) illustrates the temporal evolution of the gamma-ray flux across different timescale propagation of SNR~G296.5+10.0.

Figure~\ref{gamma_models_1}-(a) shows the predicted gamma-ray SEDs for the quiescent model of SNR~G296.5+10.0 evaluated for three spectral indices: $\alpha = 1.8$, $2.0$ and $2.2$. An important feature observed in this panel is that harder injection spectra (i.e., lower $\alpha$ values) result in a stronger high-energy gamma-ray flux, particularly above $\sim$10~GeV. This behavior reflects the increased population of high-energy protons in the source spectrum, which enhances pion production and subsequent gamma-ray emission through hadronic interactions. The differences among the curves highlight the sensitivity of the hadronic gamma-ray output to the spectral index, emphasizing the importance of spectral hardness when constraining the CR contribution of young SNRs to the observed gamma-ray flux.

\begin{figure*}[!ht]
  \centering
   \subfloat[Helium]{\includegraphics[angle=0,width=0.5\textwidth]{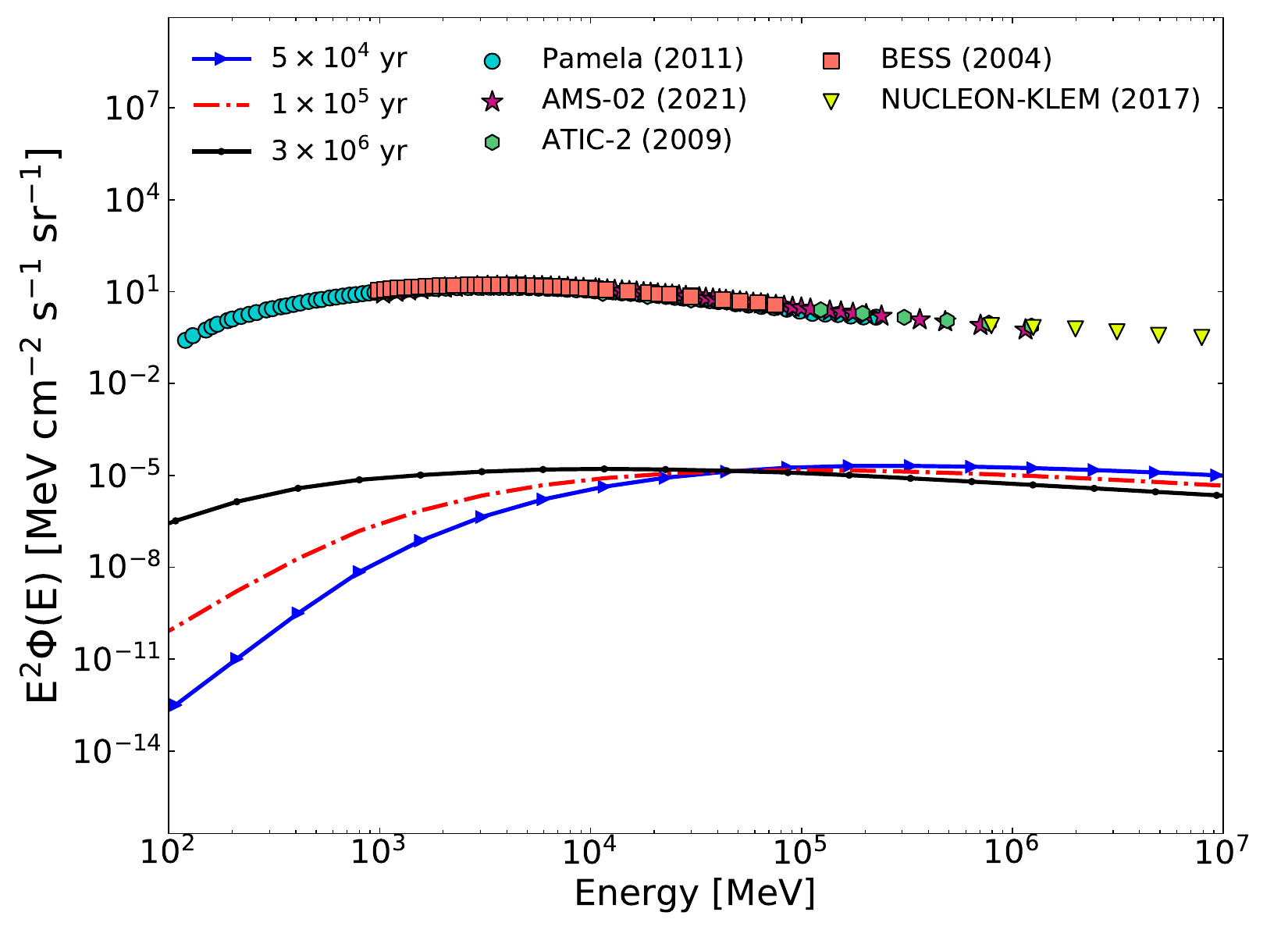}}
    \subfloat[Carbon]{\includegraphics[angle=0,width=0.5\textwidth]{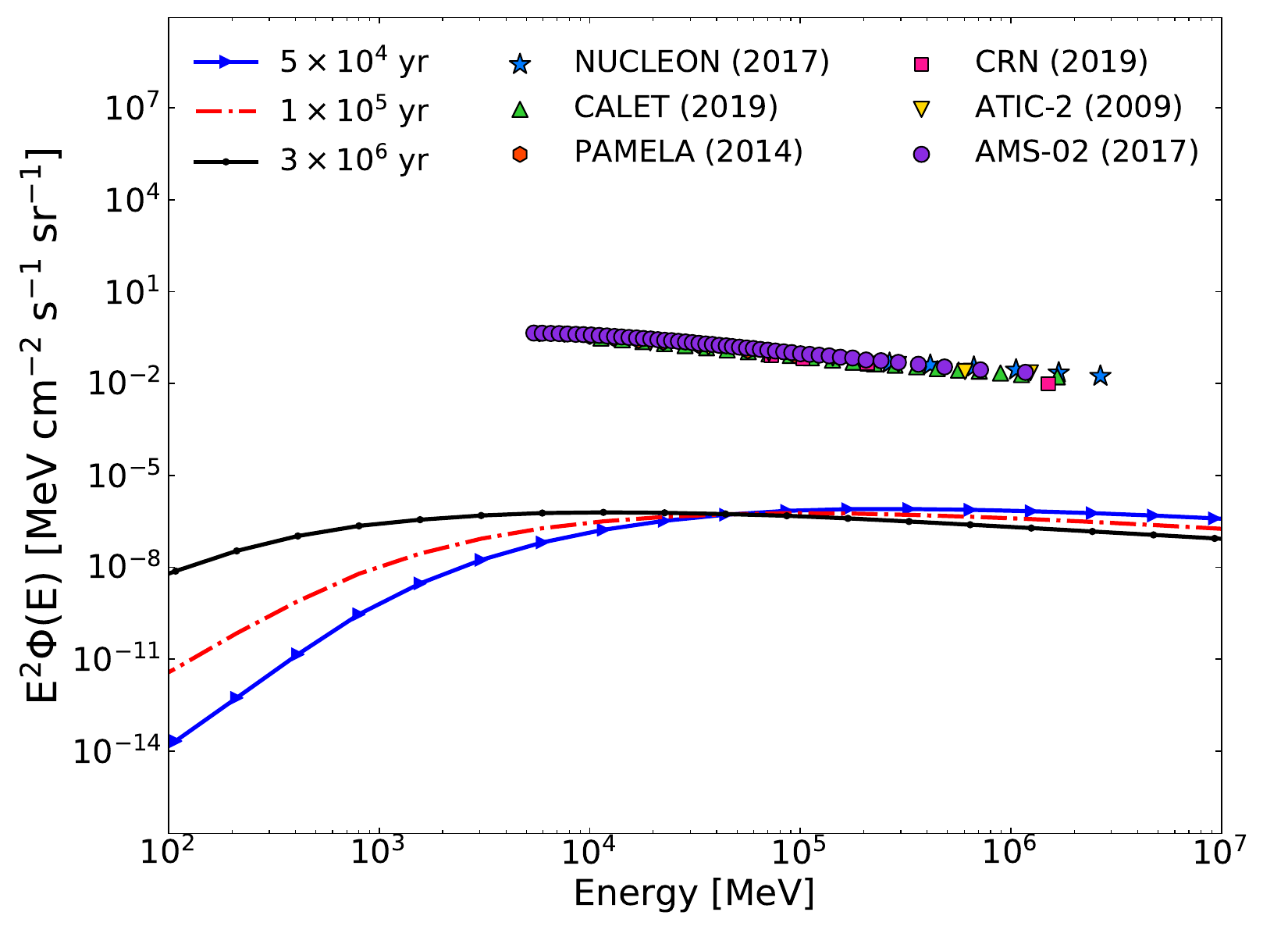}}\\
    \subfloat[Nitrogen]{\includegraphics[angle=0,width=0.5\textwidth]{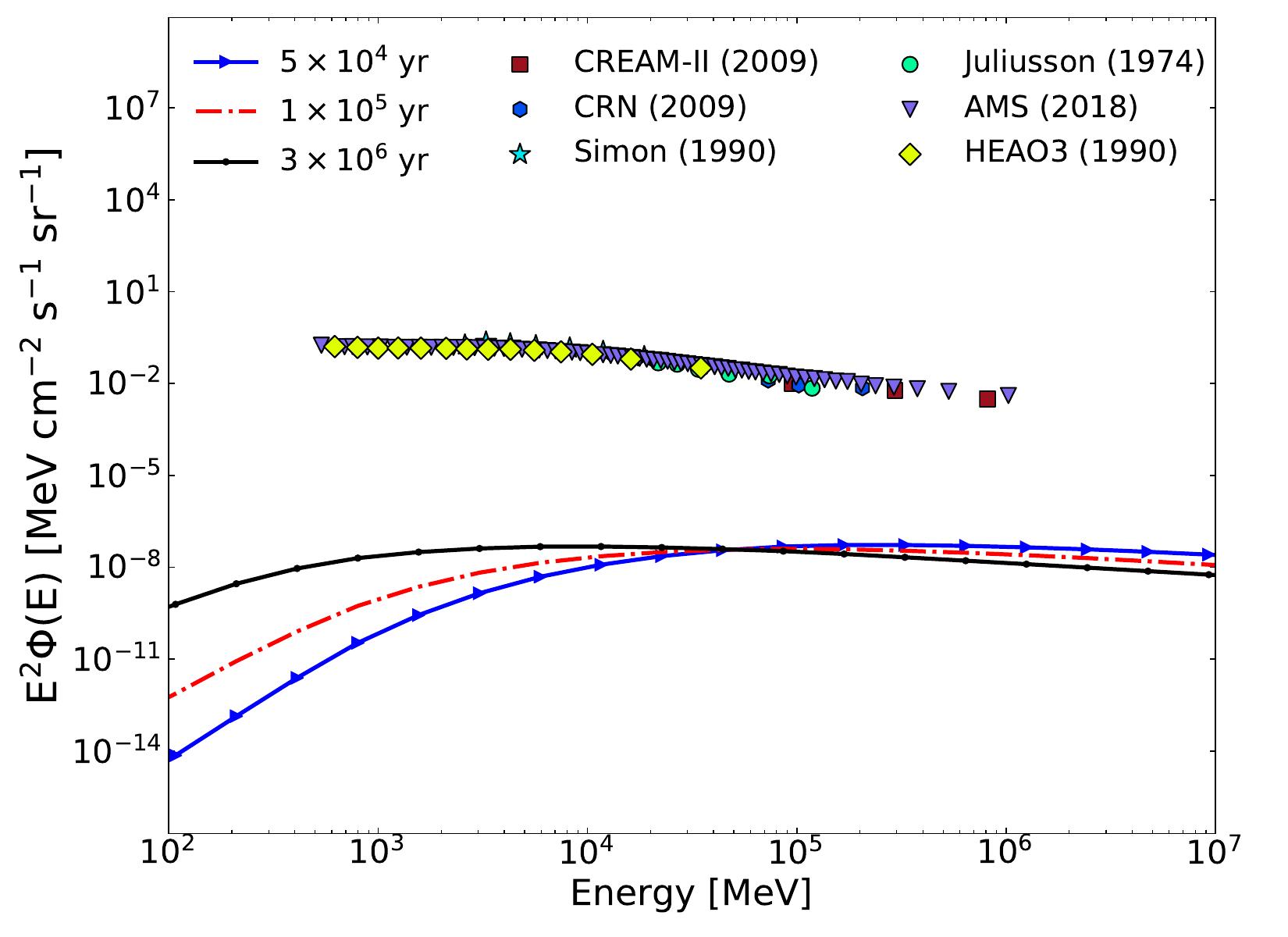}}
    \subfloat[Iron]{\includegraphics[angle=0,width=0.5\textwidth]{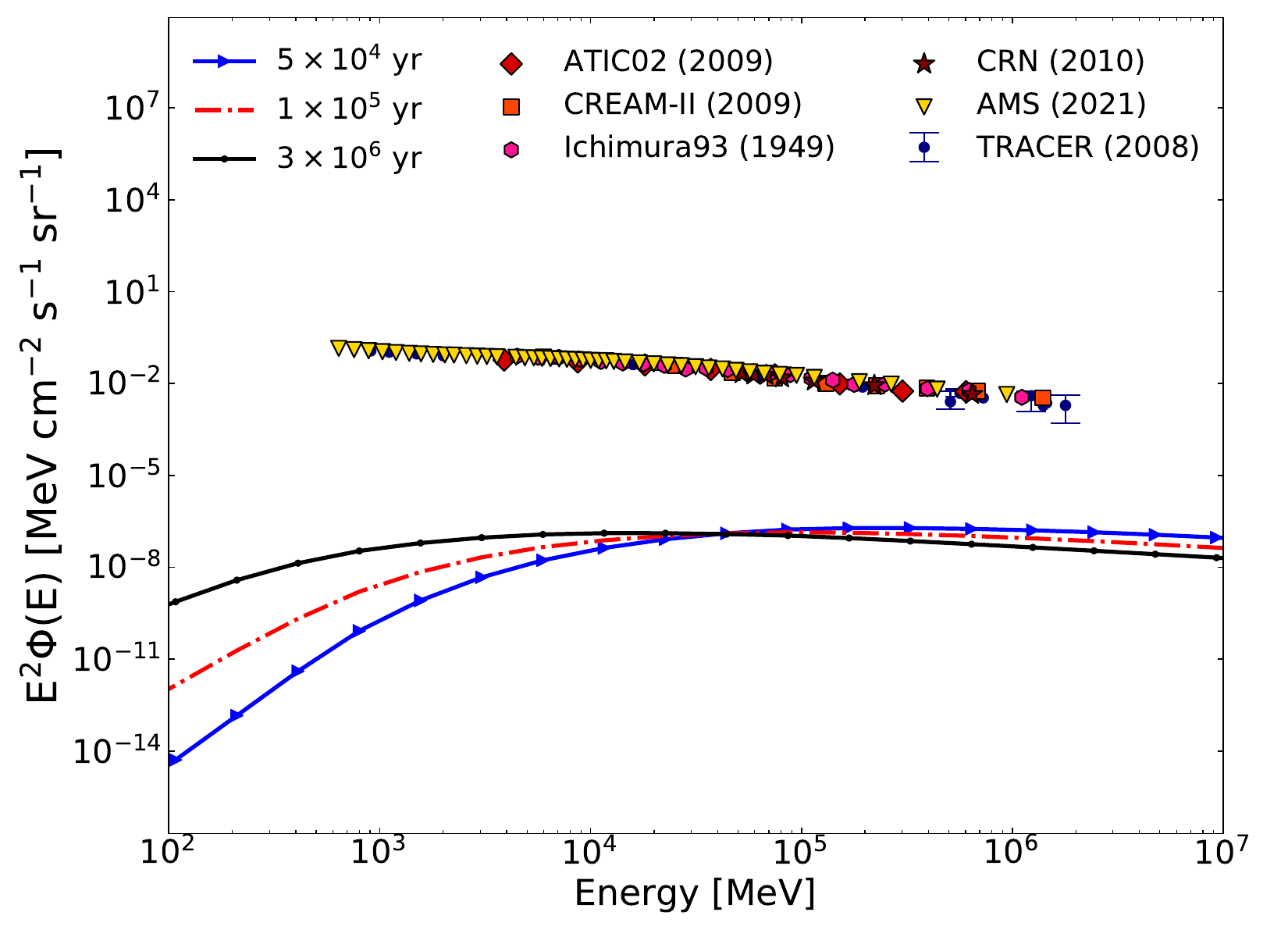}}    
    \caption{Energy spectra of helium (a), carbon (b), nitrogen (c), and iron (d) nuclei under the quiescent emission model, assuming a spectral index of $\alpha = 1.8$ and a modulation potential of 0.30 GV. The fluxes are multiplied by $E^2$ and compared with measurements from various CR observatories. The data have been extracted \cite{alemanno2021,aguilar2015he, aguilar2018ni, aguilar2021fe, adriani2020}.}
\label{spc_element}
\end{figure*}

Figure~\ref{gamma_models_1}-(b) illustrates the spin-down model, which represents the steady-state leptonic contribution from the CCO~1E 1207.4-5209. In this scenario, electrons and positrons continuously accelerated by the rotational energy loss of the CCO ($L_0 = 10^{33}$~erg~s$^{-1}$) interact with the ambient photon fields, resulting in the production of high-energy gamma rays. Although SNR dominates gamma-ray emission at intermediate energies through hadronic interactions, CCO becomes increasingly significant in the high-energy regime (above $\sim$5--10~GeV), where IC emission clearly exceeds all other components and dominates the total flux.

\begin{figure}[!ht]
    \centering
     \subfloat{\includegraphics[angle=0,width=0.50\textwidth]{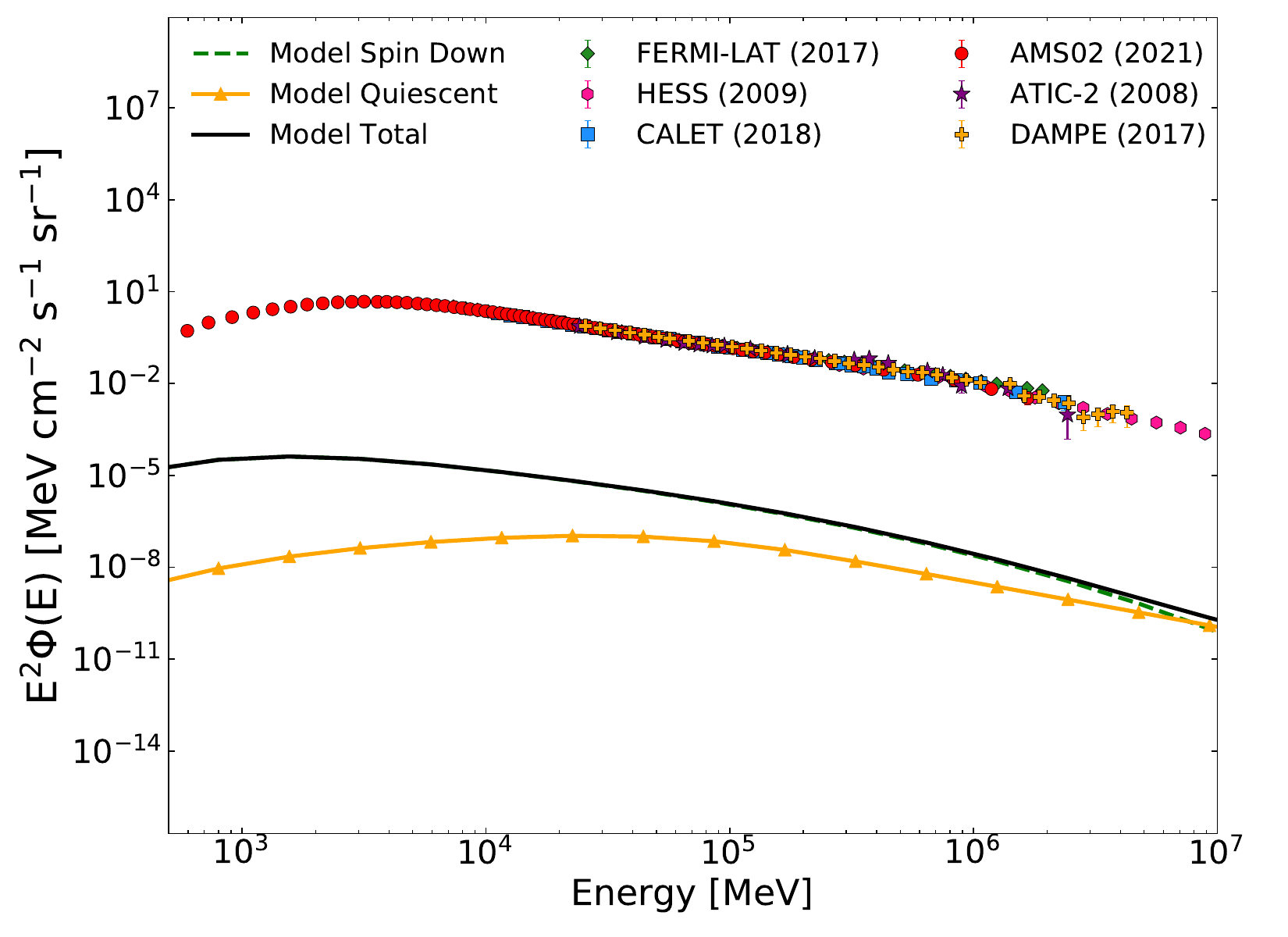}}
     %\subfloat{\includegraphics[angle=0,width=0.50\textwidth]{BoroCarbon_1.8.pdf}}
     %\caption{Left: Electron and positron energy spectra computed from the spin-down and quiescent models, corresponding to steady-state injection and a elapsed propagation timescale of $5 \times 10^{4}$ yr, respectively. The spectra are multiplied by $E^2$. The models are shown in comparison with the data considering a spectral index of $2.2$ and a modulation of 0.30 GV. The data have been extracted AMS \cite{aguilar2021}, ATIC \cite{2008Natur.456..362C}, HESS \cite{2009A&A...508..561A}, Fermi-LAT \cite{2017PhRvD..95h2007A}, DAMPE \cite{2017Natur.552...63D} and CALET \cite{2018PhRvL.120z1102A}. Right:  Boron-to-carbon (B/C) ratio as a function of kinetic energy per nucleon for the quiescent model displayed in comparison with data from \cite{adriani2014measurement,adriani2022}. This ratio provides information on CR propagation and secondary production in the interstellar medium. $B/C$ ratio for the quiescent model as a function of kinetic energy per nucleon.}
     \caption{Electron and positron energy spectra computed from the spin-down and quiescent models, corresponding to steady-state injection and a elapsed propagation timescale of $5 \times 10^{4}$ yr, respectively. The spectra are multiplied by $E^2$. The models are shown in comparison with the data considering a spectral index of $2.2$ and a modulation of 0.30 GV. The data have been extracted AMS \cite{aguilar2021}, ATIC \cite{2008Natur.456..362C}, HESS \cite{2009A&A...508..561A}, Fermi-LAT \cite{2017PhRvD..95h2007A}, DAMPE \cite{2017Natur.552...63D} and CALET \cite{2018PhRvL.120z1102A}.}
    \label{fig:BC.Electropositron}
\end{figure}

\begin{figure*}[!ht]
    \centering 
     {\includegraphics[angle=0,width=1.0\textwidth]{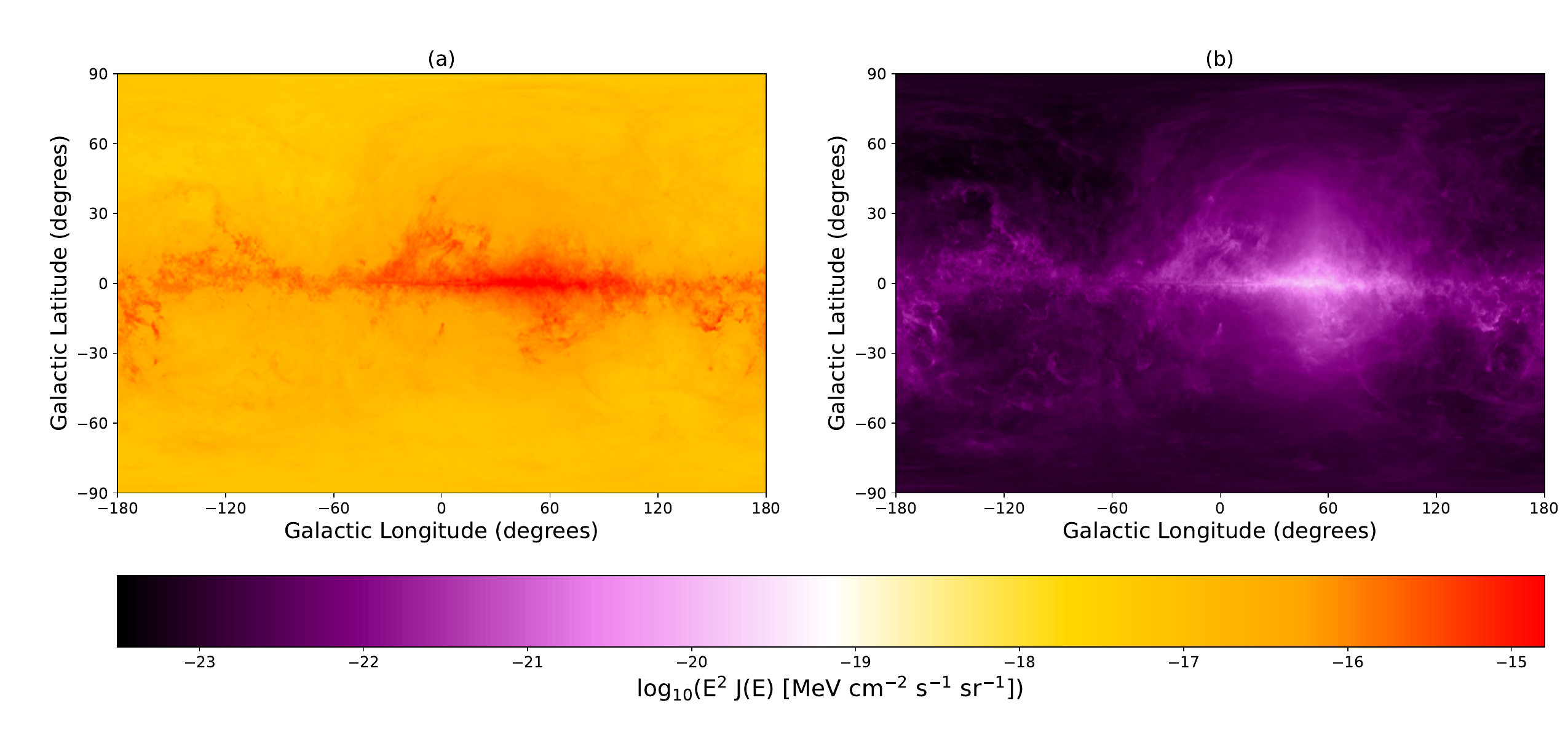}}
     \caption{Gamma-ray intensity maps at 1 GeV (a) and 0.1 TeV (b) based on the spin-down model for the CCO 1E 1207.4-5209 and its host SNR G296.5+10.0. The emission arises from bremsstrahlung processes from electrons and positrons propagated in a 3D interstellar gas density model including HI and H$_2$ distributions \cite{johannesson2018}. The maps are shown in Galactic coordinates, with (l, b) = (0, 0) at the center. The results show that the emission is spatially concentrated and correlates with regions of higher gas density, highlighting the role of the ambient medium in modulating the gamma-ray production efficiency at different energies. A faint vertical feature near $l = 0^\circ$ in panel (b) is a non-physical projection artefact.}
    \label{fig:gas}
\end{figure*}

Figure~\ref{gamma_models_2}-(a) presents the total gamma-ray SED for the quiescent model assuming a spectral index of $\alpha = 1.8$. This panel shows the individual contributions to the total flux, including hadronic gamma rays from pion decay and leptonic components such as IC scattering and bremsstrahlung. Although pion decay dominates the emission across much of the spectrum, the IC and bremsstrahlung processes introduce additional contribution, particularly at lower energies. By including these components, we can better understand how leptonic and hadronic mechanisms influence the gamma-ray production of SNR~G296.5+10.0 during its early stage, providing constraints on the source's particle acceleration environment under a hadronic-dominated scenario.

Figure~\ref{gamma_models_2}-(b) shows the evolution of the gamma-ray flux for $\alpha = 1.8$ on three elapsed propagation timescales for the quiescent model. The earliest case, $5 \times 10^4$ yr, is close to the estimated age of SNR~G296.5+10.0 and represents a realistic timescale for recent CR injection. At this stage, the quiescent hadronic component remains strongly suppressed at low energies due to the limited diffusion of protons and heavier nuclei over the $\sim$2~kpc distance to Earth. As the elapsed propagation time increases to $10^5$ and $3 \times 10^6$ yr, progressively more low-energy CR contribute to the observable flux, enhancing the hadronic signal, especially in the GeV range. At $3 \times 10^6$ yr, the system approaches a quasi-steady-state configuration, effectively setting an upper limit on the long-term contribution from this source. In particular, even at early times, the highest-energy hadronic CR diffuse more rapidly and can still contribute to the flux in the TeV range.

Together, the results in Figure~\ref{gamma_models_3} demonstrate that both hadronic and leptonic processes play an essential role in shaping the gamma-ray spectrum of the G296.5+10.0/1E~1207.4-5209 association. The steady-state leptonic emission from the CCO dominates at lower energies. In contrast, the time-evolving hadronic emission from the SNR becomes increasingly significant at higher energies. These findings support the idea that CCO can contribute to the low-energy gamma-ray flux even in the absence of a pulsar wind nebula, while the SNR shell governs the high-energy domain. Future high-sensitivity observatories like CTAO will be essential for disentangling these components and validating emission models.

We compare the propagated spectra for each nuclear species from SNR~G296.5+10.0 with the local interstellar spectra (LIS) to contextualize our gamma-ray predictions within the global CR framework. This comparison serves multiple purposes: setting conservative upper bounds on acceleration efficiency and total CR luminosity by confronting the source contribution with the LIS and gamma-ray limits \cite{2021ApJ...910...78Z,Supanitsky_2013,Anjos_2014,Sasse_2021,2021JCAP10023D,coelho2022updated}; validating the transport setup, since remaining below the LIS while preserving the observed spectral shapes for p, He, CNO, and Fe is consistent with GALPROP solutions constrained by B/C and other secondary ratios \cite{1998ApJ...509..212S,2002ApJ...565..280M,2007ARNPS..57..285S,galprop,2017JCAP...02..015E,2011ApJ...729..106T,2017A&A...597A.117D}; and verifying that fragmentation and energy-dependent escape do not introduce unphysical features in heavy nuclei for a single, distant ($\sim$2 kpc) and relatively young ($<10^5$ yr) source, a regime where discrete source effects on the LIS are expected to be small \cite{2016A&A...595A..33T}.

Figures~\ref{spc_proton} and~\ref{spc_element} show the resulting energy spectra for protons and heavier elements (He, C, N, Fe), respectively, under the quiescent emission model, revealing how the contribution from SNR~G296.5+10.0 evolves with time and varies by species. Figure~\ref{spc_proton} illustrates the energy spectra of protons, showing the source's impact on the overall CR proton flux for different spectral indices:1.8-(a), 2.0-(b), and 2.2-(c). The results indicate that for harder spectral indices (e.g., $\alpha = 1.8$), the proton flux remains substantial over a wide energy range, especially above 1~TeV. In contrast, softer spectra (e.g., $\alpha = 2.2$) result in a steeper decline in flux at higher energies. Given that the estimated age of SNR G296.5+10.0 is $10^4$~yr, the curve corresponding to $5 \times 10^4$~yr in each plot represents the current state of the remnant. At this stage, the proton flux contributes marginally to the observed CR spectrum. Even under steady state conditions, the contribution of the source remains limited, although it becomes more noticeable at lower energies, as expected due to energy-dependent propagation effects and the accumulation of low-energy particles over time.

Figure~\ref{spc_element} presents the energy spectra of heavier CR nuclei: helium (panel a), carbon (panel b), nitrogen (panel c) and iron (panel d), calculated using the quiescent emission model with a spectral index of $\alpha = 1.8$ and elapsed propagation timescales of $5 \times 10^4$, $10^5$ and $3 \times 10^6$ yr. In all panels, the modeled fluxes are shown alongside experimental data from various CR observatories. The spectra reveal a systematic decline in intensity with increasing atomic mass, reflecting the reduced abundance and enhanced fragmentation rates of heavier nuclei during their propagation through the interstellar medium. As with the proton component, the flux increases with longer propagation times, especially at lower energies, due to the delayed diffusion of cosmic rays from the SNR G296.5+10.0. However, the overall contribution of the SNR to the heavier-element fluxes remains suppressed compared to protons, reinforcing the idea that SNR G296.5+10.0 is an inefficient accelerator of high-Z nuclei. The agreement between modeled and observed spectra validates the assumptions used in the propagation model for a single-source contribution scenario.

\begin{figure*}[!ht]
   \centering
   \subfloat[Energy-dependent differential flux sensitivity and spectral model.]{\includegraphics[width=0.50\textwidth]{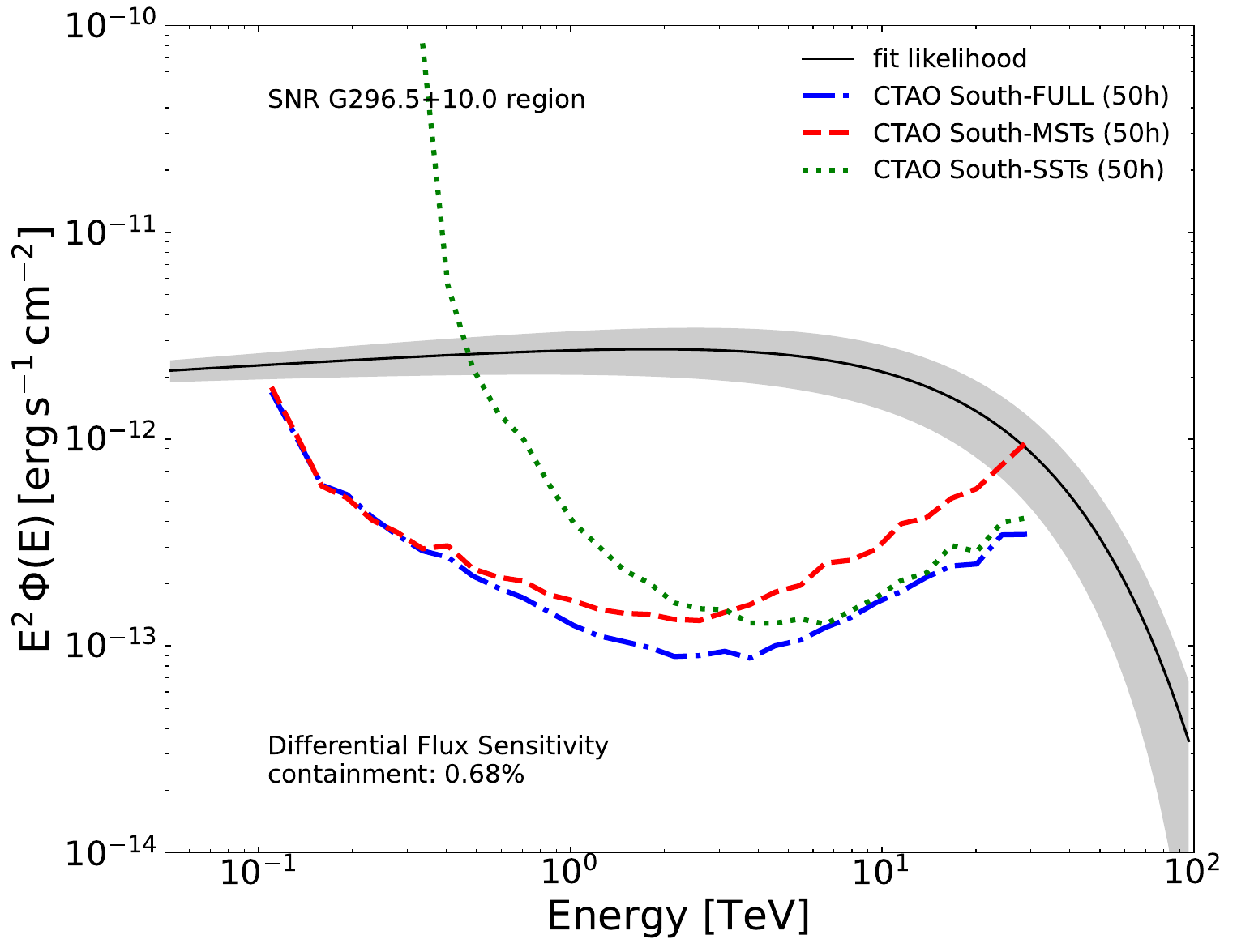}}
   \subfloat[Spectral Energy Distribution.]{\includegraphics[width=0.50\textwidth]{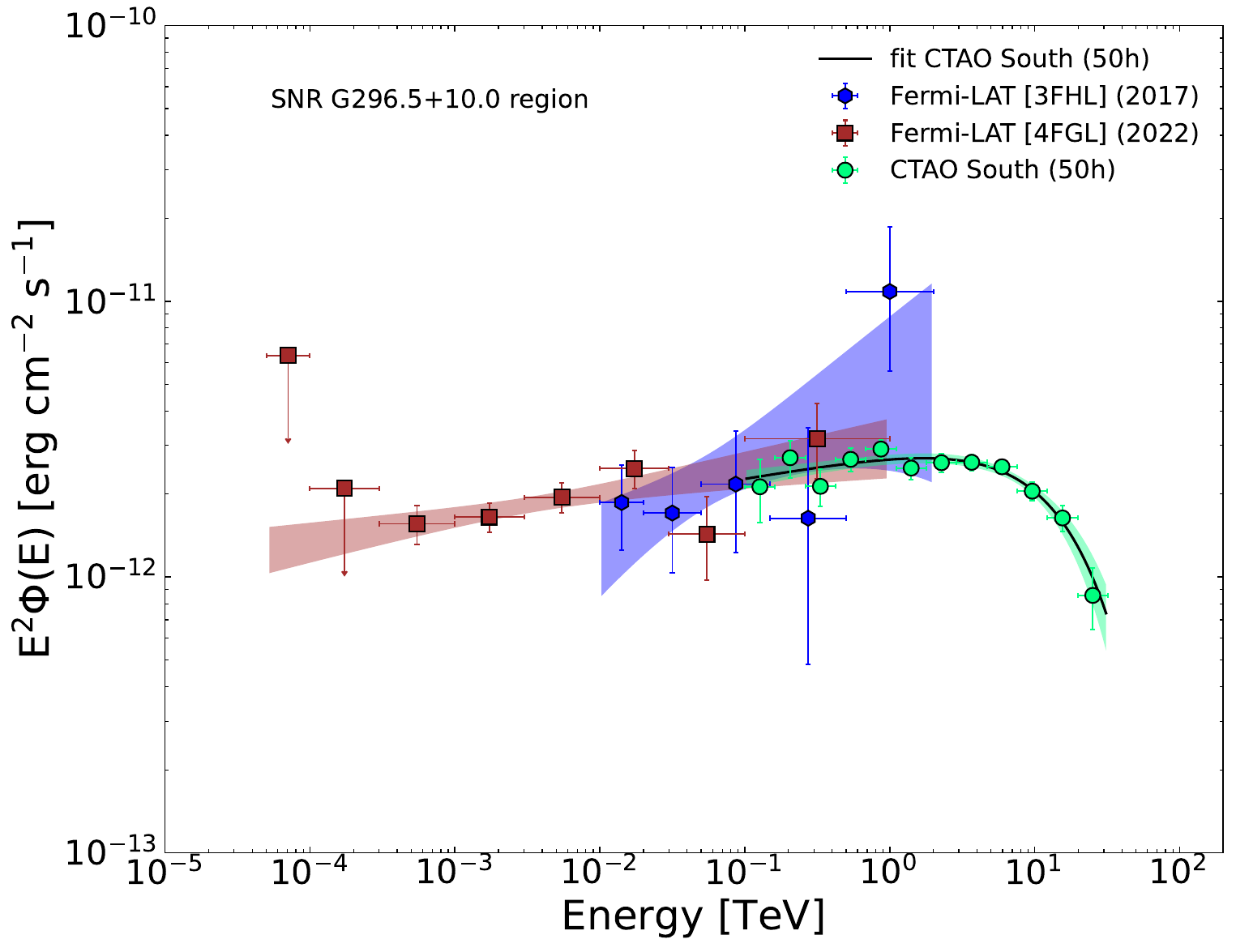}}\hfill
   \caption{Panels (a) and (b) show, respectively, the CTAO differential flux sensitivity and the SED for the SNR G296.5+10.0 region. (a) The black solid line represents the source spectral model (see Table~\ref{tab:skymodel}). Sensitivity curves for the CTAO South array configurations are displayed in different colors:(blue) CTAO South-FULL with 14 Medium-Sized Telescopes (MSTs) and 37 Small-Sized Telescopes (SSTs);(red) CTAO South with 14 MSTs only; and (green) CTAO South with 37 SSTs only. The differential flux sensitivity is computed for a 50\,h observation time using IRFs optimized for a 20$^{\circ}$ zenith angle. (b)  SED of the same region, including Fermi-LAT measurements and the CTAO South 50\,h reconstructed spectrum. The black curve corresponds to the CTAO spectral model (see Table~\ref{tab:skymodel}).  Together, the two panels illustrate the CTAO detection capabilities for weak  gamma-ray sources and the expected sensitivity improvements across the full energy range.}
    \label{gamma}
\end{figure*}

The presence of a CCO associated with the supernova remnant suggests the possibility of charged and multimessenger particle emission at low energies. However, no observational evidence currently supports the acceleration of heavy nuclei by the CCO~\cite{2000AJ....119..281G}. Figure~\ref{fig:BC.Electropositron} provides a spectrum of electrons and positrons, showing the respective contributions of the spin-down and quiescent models. The spin-down model assumes steady-state injection, whereas the quiescent model corresponds to a elapsed propagation timescale of $5 \times 10^{4}$ yr. Notably, the spin-down scenario yields a pronounced enhancement in leptonic flux at energies above $10^6$ eV, a key feature for secondary gamma-ray production through IC scattering. Together with Figures~\ref{spc_proton} and~\ref{spc_element}, which present the energy spectra for protons and heavier nuclei under different injection conditions, these results underscore the distinct roles played by the SNR and the CCO: the SNR contributes broadly to the hadronic CR population, whereas the CCO's influence is most evident in the leptonic component.

Figure~\ref{fig:gas} shows the gamma-ray intensity maps predicted by the spin-down model for the source 1E~1207.4-5209, computed with the \texttt{GALPROP} code. The maps display only the contribution from this single source, i.e., the bremsstrahlung emission from electrons and positrons injected by the CCO, propagated through the Galaxy and integrated along the line of sight according to GALPROP's built-in mapping procedure \cite{galprop}. Diffuse cosmic rays are not included in the construction of the maps around the source. Panels (a) and (b) show the intensity at 1~GeV and 0.1~TeV, respectively, using the realistic three-dimensional Galactic gas distribution from \citep{johannesson2018}. The results underline that the predicted emission from 1E~1207.4-5209 is spatially concentrated and strongly dependent on the ambient gas distribution, with the highest intensities occurring where the gas density is greatest. The source's contribution is significant only at low to medium energies; at TeV energies, its flux is too low to account for the observed high-energy gamma rays, suggesting that additional processes or sources would be required to explain any very-high-energy detection.

Finally, the left panel of Figure~\ref{gamma} presents the differential flux sensitivity curves for the CTAO South configuration, together with the spectral model of  the SNR G296.5+10.0 region (black curve), whose uncertainty is represented by the shaded band. The sensitivity curves illustrate the detection capabilities for different configurations: CTAO South-FULL (14 MSTs + 37 SSTs), CTAO South-MSTs (MSTs only), and CTAO South-SSTs (SSTs only). IRFs are optimized for a 50-hour exposure at a $20^\circ$ zenith angle with azimuth-averaged pointing. The CTAO sensitivity curves exhibit the expected behaviour: the sensitivity improves with energy up to around  $\sim 1$ TeV, reaches its minimum near this range, and then gradually worsens at higher energies. The comparison with the source model indicates that a 50-hour exposure is sufficient for CTAO to detect the gamma-ray emission across a broad energy range, particularly around the TeV scale where the instrument achieves its best performance.

The right panel of Figure~\ref{gamma} compares the modeled and observed gamma-ray fluxes in the SNR G296.5+10.0 region. The observed data points include measurements from Fermi-LAT (3FHL and 4FGL). The plotted theoretical model is fit to the data using an ECPL  spectral model (see Equation~\ref{eq:ecpl} and Table~\ref{tab:skymodel}), with a maximum cutoff energy of $E_{\mathrm{cut}} = 20$ TeV. The CTAO South (50h) curve predicts the expected sensitivity, demonstrating that CTAO will significantly enhance gamma-ray detection in this region. The agreement between the observed data, theoretical models, and CTAO expectations suggests that the SNR G296.5+10.0 region may host an efficient CR accelerator. The uncertainty bands highlight potential variations in the spectral modeling and observational constraints. The results depicted in both panels confirm that CTAO’s improved sensitivity will allow the detection of weak gamma-ray sources, helping to constrain the nature of CR interactions in the SNR G296.5 + 10.0 region.

Building on these results, we propose a two-phase observational campaign with CTAO. The first phase consists of a deep 50h exposure centered on 1E~1207.4-5209, designed to resolve the IC component with an angular resolution of $0.1^\circ$ at 1\,TeV. This will enable a spectral curvature analysis sensitive to differences of $\Delta\Gamma > 0.5$ at significance $5\sigma$, providing a decisive test for the leptonic dominance predicted by the spin-down model. The second phase involves a complementary 10h wide-field mosaic that covers the full $2^\circ$ extent of the SNR, targeting hadronic gamma-ray emission from shock-cloud interactions to flux levels of $10^{-13}$\,erg\,cm$^{-2}$\,s$^{-1}$. 

CTAO's sensitivity exceeds that of H.E.S.S. by a factor of approximately three at 1\,TeV \cite{Holler2017ICRC}, allowing for two major advances. First, it allows robust discrimination between leptonic and hadronic emission models via high-precision spectral curvature measurements \cite{2019scta.book.....C}. Second, CTAO’s superior angular resolution will allow detailed mapping of the $0.1^\circ$--$0.3^\circ$ spatial extension of the emission, effectively isolating the CCO's contribution from the surrounding SNR shell. Together, these observations will spatially and spectrally disentangle the CCO’s particle acceleration signature from the remnant’s hadronic background, while also revealing the broader CR injection geography of the system.

\section{Summary}
\label{sec:summary}

{In this study, we employed \texttt{GALPROP} in steady-state mode to model hadronic and leptonic emission from the CCO~\allowbreak1E~1207.4-5209 and its host SNR~G296.5+10.0, evaluating the predicted gamma-ray flux from CR interactions at different source ages $t_{\mathrm{age}}$ to bracket the plausible evolutionary stages of the system}. Under the spin–down scenario for the CCO, continuous injection of electrons and positrons driven by dipole radiation contributes only at low gamma–ray energies (below a few GeV), in line with the measured surface magnetic field of $B\sim8\times10^{10}\,$G. The SNR G296.5+10.0, evolving over elapsed propagation timescale from $5\times10^{4}\,$yr up to $3\times10^{6}\,$yr, produces gamma rays via hadronic and leptonic processes, whose high-energy tail hardens for injection indices $\alpha \lesssim 2.0$ but remains sub-dominant at TeV energies when compared with current upper limits. By imposing the integral flux upper limit
\begin{equation*} 
\qquad F_{\gamma}(1\,\mathrm{GeV}<E<1\,\mathrm{TeV}) \;<\; 1.2\times10^{-9}\,\mathrm{cm}^{-2}\,\mathrm{s}^{-1}
\end{equation*}
from \emph{Fermi}‐LAT, we derive an overall low CR luminosity $W_{\rm CR} \lesssim 10^{48}\,$ erg, indicating an inefficient acceleration of nuclei to very high energies. The predicted hadronic flux above 1 TeV,
\begin{equation}
\qquad \qquad F(>1\,\mathrm{TeV})\;\sim\;2.5\times10^{-14}\,\mathrm{cm}^{-2}\,\mathrm{s}^{-1}
\end{equation}
for a target density $n=0.3\,$cm$^{-3}$, lies just below the CTAO South 50 h differential sensitivity curve (Fig.~\ref{gamma}). Scaling the exposure, a $5\sigma$ detection would require on the order of 200 h of CTAO observations. The age discrepancy between the CCO ($t_{\rm CCO}\sim3.02\times10^{8}\,$yr) and the SNR ($t_{\rm SNR}\sim10^{4}\,$yr) remains unresolved \cite{2002ApJ...569L..95P, ankay2007, 2011A&A...525A.106D}. Two viable explanations are: significant fallback accretion altered the early spin-down of 1E 1207.4-5210, reducing its present dipole torque and leading our model to overestimate the current leptonic injection by $\sim$ a factor of two, and the SNR age is currently underestimated, perhaps due to interaction with a dense cloud that slows the shock. Further observations could help distinguish between these scenarios.

Overall, our results and findings emphasize the need for refined propagation models to interpret low‐luminosity gamma‐ray sources in the Galaxy. They also show that the upcoming observatory CTAO will be essential for disentangling leptonic and hadronic emission components and for probing the contribution of CCOs to CR acceleration.

\section{Acknowledgments}

We thank the referee for a thorough and constructive review that significantly improved the clarity and physical interpretation of this work. The authors acknowledge the use of computational resources from the \href{https://computacaocientifica.ufes.br/scicom}{Sci-Com Lab} of the Department of Physics at Universidade Federal do Espírito Santo (UFES), supported by Fundação de Amparo à Pesquisa e Inovação do Espírito Santo (FAPES),  Coordenação de Aperfeiçoamento de Pessoal de Nível Superior (CAPES), and Conselho Nacional de Desenvolvimento Científico e Tecnológico (CNPq). This study was financed in part by the CAPES, Finance Code 001. L.P, R. J. C, R.C.A. and J.G.C. acknowledge the financial support from the NAPI “Fenômenos Extremos do Universo” of Fundação de Apoio à Ciência, Tecnologia e Inovação do Paraná.J.G.C. is grateful for the support of FAPES (1020/2022, 1081/2022, 976/2022,
332/2023, 1514/2025), CNPq (311758/2021-5, 306018/2025-0), and FAPESP (grant
No. 2021/01089-1). R.C.A. research is supported by CAPES/Alexander von Humboldt Program (88881.800216 /2022-01), CNPq (310448/2021-2) and (4000\allowbreak045/2023-0), Araucária Foundation (698/2022) and (721/202 2) and FAPE-SP (2021/01089-1). R.C.A. gratefully acknowledges the Max Planck Institute for Nuclear Physics for their warm hospitality and support during her visit, which provided a conducive environment for fruitful discussions and collaborations. R.C.A. also acknowledges the support of L’Oreal Brazil, with the partnership of ABC and UNESCO in Brazil. Furthermore, the authors acknowledge AWS Cloud Credit/CNPq and the National Laboratory for Scientific Computing (LNCC/ MCTI, Brazil) for providing HPC resources through the SDumont supercomputer, which significantly supported the computational aspects of this research, which have contributed to the research results reported in this paper. URL: \href{https://sdumont.lncc.br}{sdumont.lncc.br}. The research also used \texttt{Gammapy.org}, a Python package developed by the community for TeV gamma-ray astronomy \cite{Deil_2017, Donath_2023}, accessible at \href{https://www.gammapy.org}{gammapy.org}. We also acknowledge the use of the Python packages \href{https://www.astropy.org/}{Astropy} \citep{astropy2013, astropy2018} and \href{https://matplotlib.org/stable/}{Matplotlib} \citep{hunter2007} .We thank the author Jorge Ariel Combi for kindly providing the gamma-ray data used in this work. In addition, we used the instrument response functions for the CTAO provided by the CTA Consortium and CTAO. For detailed information on these instrument response functions, see \href{https://www.ctao-observatory.org/science/cta-performance}{ctao-observatory.org} (version prod5 v0.1; \cite{CTAOIRFS}).

\bibliographystyle{spphys}       
\bibliography{biblio} 

\end{document}